\documentclass{aa} 
\usepackage{psfig}
\begin{document}
\def\lya{Ly$\alpha$}     %
\def\lyb{Ly$\beta$}      %
\def\lyc{Ly$\gamma$}     %
\def\lyd{Ly$\delta$}     %
\def\lye{Ly$\epsilon$}   %
\def\Wr{W_\mathrm{r}}   %
\def\Wo{W_\mathrm{obs}} %
\def\mr{m_\mathrm{R}}   %
\def\Mr{M_\mathrm{R}}   %
\def\mb{m_\mathrm{B}}   %
\def\Mb{M_\mathrm{B}}   %
\def\mi{m_\mathrm{I}}   %
\def\Mi{M_\mathrm{I}}   %
\def\mq{m_{s\mathrm 450}}%
\def\ms{m_\mathrm{702}} %
\def\mh{m_\mathrm{814}} %
\def\kms{~km~s$^{-1}$}   %
\def\cm2{~cm$^{-2}$}     %
\def\za{z_\mathrm{a}}   %
\def\zd{z_\mathrm{d}}   %
\def\ze{z_\mathrm{e}}   %
\def\zg{z_\mathrm{g}}   %
\def\h50{h_{50}^{-1}}    %
\def\hi{H\,{\sc i}}      %
\def\ci{C\,{\sc i}}      %
\def\nhi{N(\mbox{H\,{\sc i}})}%
\def\nhii{N(\mbox{H\,{\sc ii}})}%
\def\nci{N(\mbox{C\,{\sc i}})}%
\def\nsiii{N(\mbox{Si\,{\sc ii}})}%
\def\nfeii{N(\mbox{Fe\,{\sc ii}})}%
\def\nsiii{N(\mbox{Si\,{\sc ii}})}%
\def\nmnii{N(\mbox{Mn\,{\sc ii}})}%
\def\nniii{N(\mbox{Ni\,{\sc ii}})}%
\def\nznii{N(\mbox{Zn\,{\sc ii}})}%
\def\lnhi{\log(N(\mbox{H\,{\sc i}}))}%
\def\lnfeii{\log(N(\mbox{Fe\,{\sc ii}}))}%
\def\lnmgii{\log(N(\mbox{Mg\,{\sc ii}}))}%
\def\lnsiii{\log(N(\mbox{Si\,{\sc ii}}))}%
\def\lnniii{\log(N(\mbox{Ni\,{\sc ii}}))}%
\def\lncrii{\log(N(\mbox{Cr\,{\sc ii}}))}%
\def\lncaii{\log(N(\mbox{Ca\,{\sc ii}}))}%
\def\lnznii{\log(N(\mbox{Zn\,{\sc ii}}))}%
\def\lnmnii{\log(N(\mbox{Mn\,{\sc ii}}))}%
\def\alii{Al\,{\sc ii}}  %
\def\aliii{Al\,{\sc iii}}  %
\def\caii{Ca\,{\sc ii}}  %
\def\caiii{Ca\,{\sc iii}}%
\def\ci{C\,{\sc i}}    %
\def\cii{C\,{\sc ii}}    %
\def\ciii{C\,{\sc iii}}  %
\def\civ{C\,{\sc iv}}    %
\def\crii{Cr\,{\sc ii}}  %
\def\feii{Fe\,{\sc ii}}  %
\def\siii{Si\,{\sc ii}}  %
\def\siiii{Si\,{\sc iii}}%
\def\siiv{Si\,{\sc iv}}  %
\def\mgi{Mg\,{\sc i}}    %
\def\mgii{Mg\,{\sc ii}}  %
\def\oi{O\,{\sc i}}      %
\def\ni{N\,{\sc i}}      %
\def\ovi{O\,{\sc vi}}    %
\def\nv{N\,{\sc v}}      %
\def\znii{Zn\,{\sc ii}}  %
\def\niii{Ni\,{\sc ii}}  %
\def\crii{Cr\,{\sc ii}}  %
\def\mnii{Mn\,{\sc ii}}  %
\thesaurus {03(11.17.1; 11.08.1; 11.09.4)} 
\title{A HST Spectroscopic study of QSOs with intermediate redshift damped 
\lya\ systems\thanks{Based on observations made with the NASA/ESA {\it Hubble Space 
Telescope}, obtained at the
Space Telescope Science Institute, which is operated by the Association of 
Universities for
Research in Astronomy, Inc., under NASA contract NAS 5-26555}
}  
\author{Patrick Boiss\'e\inst{1}, Vincent Le Brun\inst{2}, Jacqueline 
Bergeron\inst{3,4}, Jean-Michel Deharveng\inst{2}}   
\institute{
Ecole Normale Sup\'erieure, 24 rue Lhomond, F-75005 Paris, France, 
boisse@ensapa.ens.fr
\and Laboratoire d'Astronomie Spatiale du C.N.R.S., B.P. 8, F-13376 
Marseille, France, vlebrun@astrsp-mrs.fr, jmd@astrsp-mrs.fr
\and European Southern Observatory, Karl-Schwarzschild-Stra\ss e 2., D-85748
Garching b. M\" unchen, Germany, jbergero@eso.org
\and Institut d'Astrophysique de Paris, CNRS, 98bis boulevard Arago, 
F-75014 Paris, France}
\offprints{P. Boiss\'e}
\date {Received 29 August 1997 ; accepted 6 January 1998}
\maketitle
\markboth{}{}
\maketitle        
\begin{abstract} 
We present HST spectra for a sample of six QSOs with intermediate redshift
($\za\le 1.$) damped \lya\ systems. These observations aim at measuring the 
\hi\ column density and detect metal lines in order to investigate the metal
enrichment of the gas, as well as the presence of neutral species, molecules
and dust.

All systems selected on the basis of 21~cm absorption and/or strong \feii\  
lines relative to \mgii\ ones
turn out to have $\nhi$ larger than $10^{20}$\cm2. From our detection of
weak lines from minor metals and already published optical data, we determine
relative abundances of Si, Mn, Fe, Ni, Zn. In PKS~1229$-$021, we measure
[Zn/H] $= -0.5$ at $\za = 0.3950$ while in two other cases 
with intervening spiral galaxies and for which
only [Fe/H] and [Mn/H] could be estimated, the metallicity
could be close to solar. Thus, it appears that although
the scatter of metallicities is large at $\za\le 1.$ as at high redshift, 
an increasing proportion of systems with metallicities $\simeq$ 30\% solar 
are found when going at lower redshifts.

\ci\ lines are tentatively detected in two systems. Given the low metallicity,
the observed \ci/\hi\ ratio suggests that physical conditions in the absorbers 
are
comparable to those in our Galaxy. In PKS~1229$-$021, the 21~cm absorption data 
combined with the new \lya\ observations, imply a low temperature, $T_\mathrm{s} \le
200$~K, for the 
$\za = 0.3950$ absorbing gas. For the three systems in which they could 
be searched for, H$_2$ molecules are not detected with an
upper limit of about $10^{18}$\cm2\ on $N(\mathrm{H}_2)$. No evidence
is found for Galactic-type dust, except possibly in the 3C~286
$\za = 0.6922$ system. 

Our results suggest that available observations may be biased against 
dust-rich absorbers. Further, when all available measurements 
of $\nhi$ and [Zn/H] are considered,
a clear deficiency of systems with large $\nhi$ {\it and} high metallicity is
apparent. We conclude that dust extinction causes a 
preferential selection of QSOs with intervening gas relatively poor in metals,
dust and molecules. As a consequence, the high end of the \hi\ column 
density distribution (and hence $\Omega_\mathrm{g}$, the contribution of neutral 
gas to the cosmological mass density)
is probably more heavily underestimated than previously thought, 
especially at low redshift. Such a bias could also explain the high incidence of 
non-spiral morphologies in our sample.

We stress that observation of a larger sample of low $z$ damped \lya\ systems
as well as surveys of damped \lya\ systems in fainter QSOs
would give a more representative view of the true 
diversity of absorber properties and should help to probe the denser 
phases of the interstellar medium in distant galaxies.

Our program also provides a few new results on other (likely
non-damped) metal systems. In PKS~0454+039, we detect for the first time \mgii\ 
absorption from a dwarf galaxy halo at $\za = 0.072$. Strong \mgii\ 
and \feii\  absorption is also found in EX~0302$-$223 
from a spiral galaxy at $z = 0.118$. 

\end{abstract}   
%
\section{\label{intro}Introduction}
The study of high redshift damped \lya\ absorption systems (hereafter DLAS)
detected in the 
optical spectra of QSOs is now recognized as a powerful means to investigate 
the 
properties of distant galaxies or protogalaxies. Many characteristics of the 
associated 
intervening objects can be determined from absorption line data: \hi\ column 
density, metallicity, ionization state and velocity distribution of the gas. 
For radio-loud quasars, even more physical parameters are accessible 
such as the temperature of the \hi\ gas (if 21~cm absorption has been searched 
for) or 
Faraday rotation measures induced by the intervening plasma. The large $\nhi$ 
values together with the low ionization degree and small velocity dispersion 
inferred from ground-based observations for the $z \ge 1.7$ absorbers 
bear much resemblance to the properties observed in
local galactic disks (Wolfe et al. 1986). Further, in average, metals are 
observed to be underabundant with respect to Solar values but by a factor 
of typically 10 (Pettini et al. 1997a) which is naturally accounted 
for by cosmic evolution 
(Pei \& Fall 1995). DLAS are therefore believed 
to trace the internal regions of distant "normal" galaxies or their progenitors 
up to the redshifts of the most distant QSOs known to date ($\simeq 4.9$). 
\begin{table*}
\caption{\label{obslog} QSO/Damped \lya\ system characteristics and observation 
log}
\smallskip
\begin{tabular}{llrrlrlll}
\hline\noalign{\smallskip}
Quasar       & \multicolumn{2}{c}{Coordinates (J2000)} & $\ze$ & $\za$     & 
M$_\mathrm{AB}$(B)& Grism  & Exposure                   & Observation \\
             & R.A.           & Dec                    &       &           &     
  &      &time  
                                &    date          \\
\hline\noalign{\smallskip}
EX 0302$-$223   & 03 02 36.1 & -22 23 34 & 1.409 & 1.0095 (DLAS cand.)     & 
$-$19.60  
& G270H & 2400~s          & 07 Dec 1995 \\
PKS 0454+039 & 04 56 47.1 & +04 00 53 & 1.345 &  0.8596 (DLAS)  & $-$19.95   & 
G190H &
 3640~s          &  3 Jan 1995 \\
             &            &           &       &                            & 
&G270H &
$1500$~s                        &             \\
3C 196       & 08 13 36.0 & +48 13 03 & 0.871 & 0.437 (DLAS, 21~cm) & $-$21.52 & 
G160L &
3200~s          & 14 Oct 1994 \\
Q   1209-107 & 12 11 40.6 & +10 30 03 & 2.191 & 0.6295 (DLAS cand.) & $-$21.21 & 
G160L &
3200~s             &  3 Feb 1995 \\
PKS 1229$-$021 & 12 32 00.0 &$-$02 24 05& 1.038 & 0.3950 (DLAS, 21 cm)  & 
$-$18.42
& G190H & 3640~s             &  1 Jan 1995 \\
             &            &           &       &                            & 
&G270H &
$1600$~s                         &             \\
3C 286       & 13 31 08.3 & +30 30 32 & 0.849 & 0.6922 (DLAS, 21~cm)  & $-$20.17 
& G190H & 3870~s             & 26 Jan 1995 \\
             &            &           &       &                            & 
&G270H &
$1600$~s                         &             \\
\noalign{\medskip}\hline
\end{tabular}
\end{table*}

To verify this assumption in a direct way, and especially investigate the
contribution of faint objects (such as 
dwarf or low surface-brightness galaxies) that QSO absorption line 
studies have the power to reveal, we have undertaken a HST study of 
damped 
(or candidate damped) \lya\ systems at low $z$. Indeed, because of their anticipated 
faintness
and low angular separation to the QSO, it is very difficult to detect and 
characterize the associated intervening objects using images obtained from the 
ground 
(see e.g. Boiss\'e \& Boulade 1990; Steidel et al. 1994a).
In the latter, the presence of normal galaxies can be detected at low impact 
parameters but their morphological type, magnitude or extent can hardly be 
determined due to blending with the bright QSO image. Fainter objects or 
galaxies at angular 
separations smaller than about 0.5\arcsec\ cannot be reliably detected.

In a previous paper (Le Brun et al. 1997, hereafter Paper I), we used high 
angular resolution HST images to search for galaxies close to the line of sight
of QSOs with $z \le 1$ damped (or candidate damped) \lya\ systems.
Thanks to a careful subtraction of the QSO image, we could show that galaxy-like
objects are always present at small impact parameters and that, contrary to 
\mgii\ absorbers, damped \lya\ lines are associated with objects displaying a 
broad range of morphologies and surface brightness.

The present paper is devoted to a spectroscopic study of the same
QSO sample. Our primary aim is to complement data already obtained on a few 
intermediate redshift systems (Steidel et al. 1993; Cohen et al. 1994; 
Cohen et al. 1996) and 
investigate several other cases that have not yet been studied. More 
specifically we wish i) to determine the \hi\ column density 
and confirm the damped nature of each system, ii) to detect metal lines suitable
for measuring the relative abundance of heavy elements and iii) to investigate 
other properties 
such as the amount of dust and molecules. Since we now have some information
on the morphological type,
luminosity and extent of the intervening galaxies and know which part of them is 
probed 
by the QSO sightline, QSO spectra can give important clues 
to connect the absorption line data to our knowledge of nearby galaxies. More 
generally, 
such a global study of systems at intermediate redshift (i.e. intermediate 
look-back time) is essential to better understand the cosmic evolution of the 
damped \lya\ absorbers from $z \simeq 0$ up to $z \simeq 4$ (Pettini et al. 
1997a). 

This paper is organized as follows: in Sect. 2, we present the observations and 
data analysis. The main new results emerging from the spectra are summarized for 
each 
target in Sect. 3 while in Sect. 4, we discuss the properties of the 
DLAS that can be inferred from the new FOS data. Some other metal-rich  
systems
of interest are presented in Sect. 5. Finally, we discuss the implications of 
these
observations on the cosmic evolution of damped \lya\ absorbers and the bias 
induced 
by dust extinction on the apparent properties of the large $\nhi$ systems.
\begin{figure*}
\centerline{\psfig{figure=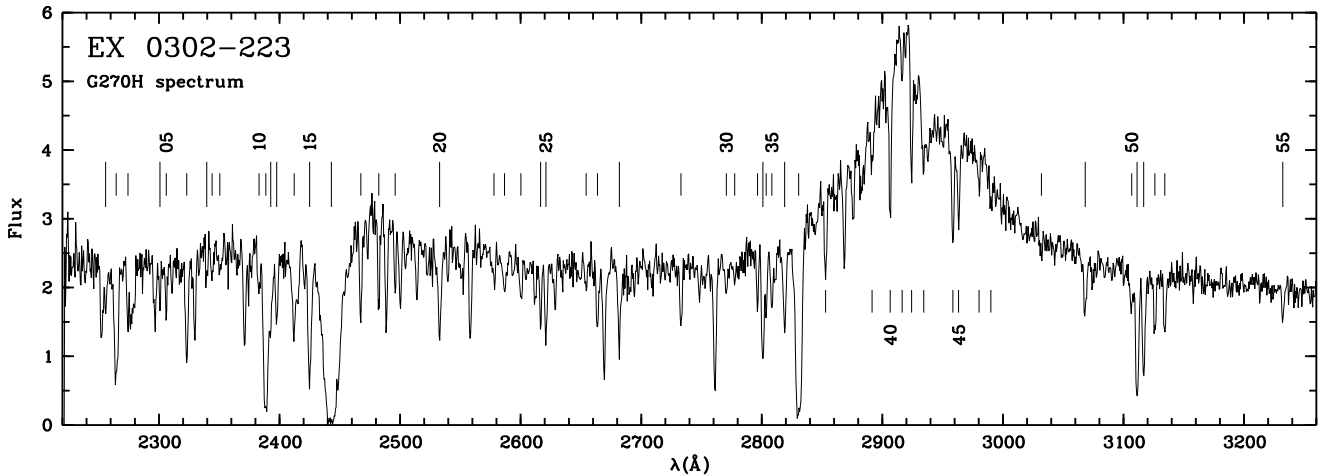,width=18cm,angle=-90,clip=true}}
\caption{\label{spec0302} FOS spectrum of EX~0302$-$223 (G270H). 
Long tick marks correspond to absorption lines from the DLAS}
\end{figure*}
\begin{table}[h!]
\def\baselinestretch{1.}
\caption{\label{q0302list} Absorption lines detected in the spectrum of 
EX0302$-$223}
\smallskip
\scriptsize{
\begin{tabular}{rrrrrlr}
N & $\lambda _\mathrm{obs}$ & $\Wo$ & $\Wr$ & $\lambda _\mathrm{r}$
& Ident. & $\za$ \\
\hline\noalign{\smallskip}
  1 & 2255.80 & 0.45 & 0.22 & 1122.53 & Fe\,{\sc iii} & 1.0096 \\ 
  2 & 2264.54 & 2.60 & 1.12 &  972.54 & \lyc          & 1.3285 \\ 
  3 & 2274.29 & 0.80 & 0.34 &  977.02 & C\,{\sc iii}  & 1.3278$^\mathrm{a}$\\ 
  4 & 2300.83 & 0.37 & 0.18 & 1144.94 & Fe\,{\sc ii}  & 1.0096 \\ 
  5 & 2306.01 & 0.51 & 0.27 & 1206.50 & Si\,{\sc iii} & 0.9113 \\
    &         &      & 0.23 & 1025.72 & \lyb          & 1.2482 \\
  6 & 2323.09 & 1.60 & 0.84 & 1215.67 & \lya          & 0.9110 \\
  7 & 2339.62 & 0.33 & 0.14 &  972.54 & \lyc          & 1.0093 \\
  8 & 2344.10 & 0.42 & 0.42 & 2344.21 & Fe\,{\sc ii}  &$-$0.0000 \\
  9 & 2350.46 & 0.49 & 0.20 &  977.02 & C\,{\sc iii}  & 1.4057 \\
 10 & 2383.05 & 0.66 & 0.66 & 2382.76 & Fe\,{\sc ii}  & 0.0001 \\
 11 & 2388.63 & 5.31 & 2.29 & 1025.72 & \lyb          & 1.3287 \\
 12 & 2391.89 & 0.62 & 0.31 & 1190.41 & Si\,{\sc ii}  & 1.0093 \\
 13 & 2397.65 & 0.86 & 0.43 & 1193.29 & Si\,{\sc ii}  & 1.0093 \\
 14 & 2412.19 & 1.46 & 0.73 & 1200.0  & N\,{\sc i}    & 1.0102$^\mathrm{a}$\\
    &         &      &      & 1036.34 & C\,{\sc ii}   & 1.3276 \\      
 15 & 2425.02 & 2.48 & 1.23 & 1206.50 & Si\,{\sc iii} & 1.0100 \\
 16 & 2442.96 & 21.5 & 10.7 & 1215.67 & \lya          & 1.0096 \\
 17 & 2467.40 & 1.03 & 0.43 & 1025.72 & \lyb          & 1.4055 \\
 18 & 2482.31 & 0.93 & 0.39 & 1031.93 & O\,{\sc vi}   & 1.4055 \\
 19 & 2495.89 & 0.54 & 0.22 & 1037.62 & O\,{\sc vi}   & 1.4054 \\
 20 & 2532.81 & 1.55 & 0.77 & 1260.42 & Si\,{\sc ii}  & 1.0095 \\
    &         &   e   &      & 1259.52 & S\,{\sc ii}  & 1.0095 \\
 21 & 2577.86 & 0.41 & 0.41 & 2576.88 & Mn\,{\sc ii}  & 0.0004 \\
 22 & 2586.66 & 0.58 & 0.58 & 2586.65 & Fe\,{\sc ii}  & 0.0000 \\
 23 & 2600.25 & 0.55 & 0.55 & 2600.17 & Fe\,{\sc ii}  & 0.0000 \\
 24 & 2616.59 & 0.66 & 0.33 & 1302.17 & O\,{\sc i}    & 1.0094 \\
 25 & 2620.96 & 0.98 &      & 2344.21 & Fe\,{\sc ii}  & 0.1181$^\mathrm{a}$\\
    &         &      &      & 1304.37 & Si\,{\sc ii}  & 1.0094 \\
 26 & 2654.27 & 0.45 & 0.40 & 2374.46 & Fe\,{\sc ii}  & 0.1178 \\ 
 27 & 2663.71 & 0.28 & 0.25 & 2382.76 & Fe\,{\sc ii}  & 0.1179 \\ 
 28 & 2681.83 & 1.53 & 0.76 & 1334.53 & C\,{\sc ii}   & 1.0096$^\mathrm{a}$\\ 
    &         &      &      & 1335.71 & C\,{\sc ii}*  & 1.0096 \\
 29 & 2732.97 & 0.77 & 0.34 & 1215.67 & \lya          & 1.2481$^\mathrm{a}$\\ 
 30 & 2770.58 & 0.37 & 0.16 & 1190.42 & Si\,{\sc ii}  & 1.3274 \\ 
 31 & 2777.51 & 0.19 & 0.08 & 1193.29 & Si\,{\sc ii}  & 1.3276 \\ 
 32 & 2796.45 & 0.61 & 0.61 & 2796.35 & Mg\,{\sc ii}  & 0.0000 \\ 
 33 & 2800.88 & 1.75 & 0.87 & 1393.76 & Si\,{\sc iv}  & 1.0096 \\ 
 34 & 2803.69 & 0.56 & 0.53 & 2803.53 & Mg\,{\sc ii}  & 0.0001 \\ 
 35 & 2808.37 & 0.84 & 0.36 & 1206.50 & Si\,{\sc iii} & 1.3277 \\ 
 36 & 2818.98 & 1.40 & 0.70 & 1402.77 & Si\,{\sc iv}  & 1.0096 \\ 
 37 & 2830.52 & 6.14 & 2.64 & 1215.67 & \lya          & 1.3284$^\mathrm{a}$\\ 
 38 & 2852.95 & 0.78 & 0.78 & 2852.96 & Fe\,{\sc ii}  &$-$0.0000\\ 
 39 & 2891.47 & 0.38 & 0.34 & 2586.65 & Fe\,{\sc ii}  & 0.1178 \\ 
 40 & 2906.47 & 1.18 & 1.06 & 2600.17 & Fe\,{\sc ii}  & 0.1178 \\ 
 41 & 2916.39 & 0.24 & 0.13 & 1526.71 & Si\,{\sc ii}  & 0.9102 \\ 
 42 & 2924.26 & 0.81 & 0.50 & 1215.67 & \lya          & 1.4055 \\ 
 43 & 2934.31 & 0.61 & 0.50 & 1260.42 & Si\,{\sc ii}  & 1.3280$^\mathrm{a}$\\ 
 44 & 2958.49 & 0.96 & 0.50 & 1548.19 & C\,{\sc iv}   & 0.9109 \\ 
 45 & 2963.26 & 0.85 & 0.44 & 1550.77 & C\,{\sc iv}   & 0.9108 \\ 
 46 & 2980.19 & 0.26 & 0.11 & 1238.82 & N\,{\sc v}    & 1.4057 \\ 
 47 & 2990.01 & 0.23 & 0.09 & 1242.80 & N\,{\sc v}    & 1.4059 \\ 
 48 & 3031.91 & 0.21 & 0.09 & 1260.42 & Si\,{\sc ii}  & 1.4055 \\ 
    &         &      & 0.09 & 1302.17 & O\,{\sc i}    & 1.3284 \\ 
 49 & 3068.21 & 1.19 & 0.59 & 1526.71 & Si\,{\sc ii}  & 1.0097 \\ 
 50 & 3106.83 & 0.73 & 0.31 & 1334.53 & C\,{\sc ii}   & 1.3280 \\ 
 51 & 3111.32 & 2.58 & 1.28 & 1548.19 & C\,{\sc iv}   & 1.0096 \\ 
 52 & 3116.74 & 2.01 & 1.00 & 1550.77 & C\,{\sc iv}   & 1.0098 \\ 
 53 & 3126.04 & 0.78 & 0.70 & 2796.35 & Mg\,{\sc ii}  & 0.1179 \\ 
 54 & 3134.25 & 0.76 & 0.68 & 2803.53 & Mg\,{\sc ii}  & 0.1180 \\ 
 55 & 3232.15 & 0.52 & 0.26 & 1608.45 & Fe\,{\sc ii}  & 1.0095 \\ 
\noalign{\medskip}\hline
\end{tabular}
\smallskip\par
$^\mathrm{a}$ Blend
}
\def\baselinestretch{2.}
\end{table}
\section{Observations and data reduction}
\subsection{The sample}
Since no confirmed DLAS had been discovered by the
HST at the time when our study was undertaken, we had to rely on
the presence of 21~cm absorption or a high
$\Wr(\mbox{\feii {$\lambda$}2382 or {$\lambda$}2600})/\Wr(\mbox{\mgii {$\lambda$}2796})$ 
ratio to select our targets 
(see Paper I for details). Three quasars in our sample satisfy the first 
criterion: 3C~196, PKS~1229$-$021 and 3C~286. It is to be noted that the 
presence of 21~cm absorption does not necessarily guarantee that the 
\lya\ line is damped 
because part of the radio flux originates from extended components such as jets 
or lobes (3C~196 and  PKS~1229$-$021) and the \hi\ distribution in the 
absorber may exhibit spatial structure at scales comparable to the  
extent of these features. The other three objects, EX~0302$-$223, 
PKS~0454+039 and Q~1209+107, have a system with a high 
$\Wr(\mbox{\feii {$\lambda$}2382 or {$\lambda$}2600})/
\Wr(\mbox{\mgii {$\lambda$}2796})$ ratio.

We have retrieved from the HST database the G270H spectrum of EX~0302$-$223 
which has been observed by another team (program 6224, dataset Y2SH0103T). 
The main characteristics of the QSOs and absorption systems of interest, 
as well as the log of the HST observations, 
are given in Table~\ref{obslog} (the absolute magnitudes of the candidate
absorbers listed in paper I are affected by an error; corrected values
are given in Table~\ref{obslog}).
\subsection{Data reduction}
The only modifications introduced to the HST pipe-line reduced data involves 
the wavelength scale and zero point of the intensity scale. When two distinct 
exposures were obtained for a given object and grism, the two spectra were 
averaged (with weights according to the exposure time). To define the absolute 
wavelength scale, we apply uniform shifts 
determined using strong Galactic lines from singly ionized species, the latter 
being 
assumed at rest. In some spectra, no such line is seen with a good enough 
signal-to-noise ratio; we then rely on strong unblended lines from already known 
absorption 
systems with well determined optical redshifts. Further, when a common 
transition is 
detected both in the G190H and G270H spectra, 
this feature is used to constrain the relative shift of these two spectra. 
The shifts applied to the original wavelengths for PKS~0454+039, PKS~1229$-$021 
and 3C~286 are 1.4, 2.3 and 1.2~\AA\ respectively for the G190H spectra 
and 1.5, 2.3 and 1.8~\AA\ for the G270H data (the G270H spectrum of 
EX~0302$-$223 has been shifted by $-0.6$~\AA\ ). 
For the G160L data, only the strongest absorption features from known systems
are useful; $-5.5$~\AA\ and 6~\AA\ have been added to the original wavelengths 
for Q~1209+107 and 3C~196 respectively. 

Regarding the intensity scale, we used the profiles of damped \lya\ lines or of 
Lyman
edges to determine the true zero level. Generally, damped \lya\ lines seen in 
the 
high resolution data do not go exactly to zero, even when they are clearly 
saturated
at the line core. Although the observed offsets are quite small, they would 
have a significant effect on the fitting procedures, so we subtracted them. 
These offsets are generally positive 
and are presumably due to scattered light in the instrument. Their 
effect on $W_\mathrm{obs}$ measurements is however always
negligible as compared to $\sigma(W_\mathrm{obs})$.
\subsection{Data analysis and results}
\begin{figure*}
\centerline{\psfig{figure=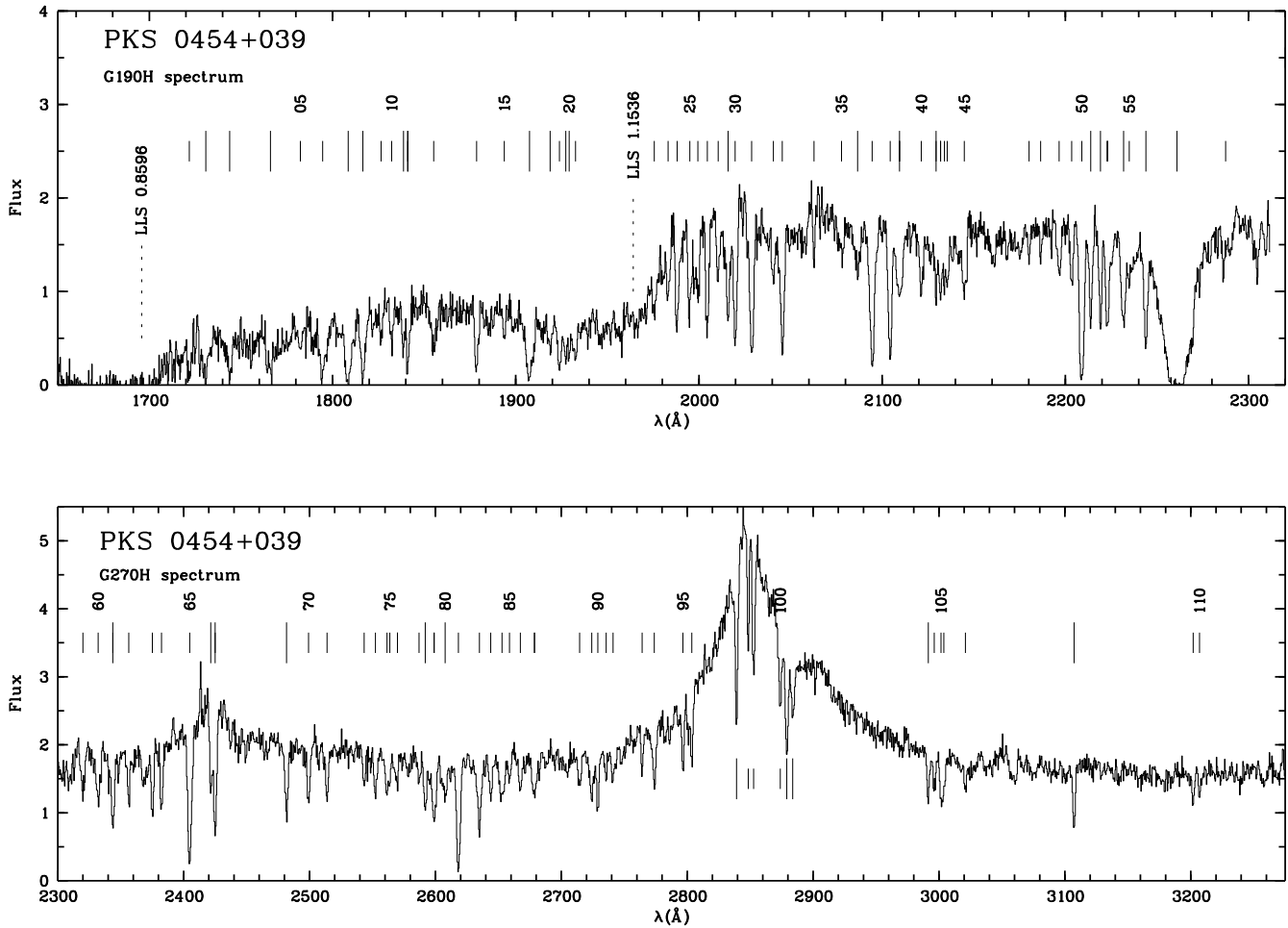,width=18cm,angle=-90,clip=true}}
\caption{\label{spec0454} Same as Fig.~\ref{spec0302} for PKS~0454+039 (G190H 
and G270H spectra)}
\end{figure*}

Figures \ref{spec0302} to \ref{spec3c286} present the spectra obtained 
for EX~0302$-$223 (G270H), PKS~0454+039 (G190H and G270H), 
3C~196 and Q~1209+107 (G160L), PKS~1229$-$021 and 3C~286 (G190H and 
G270H). The full width at half-maximum (FWHM) of an unresolved line is 1.5, 2.0 
and
6.3~\AA\ in the G190H, G270H and G160L data respectively. All spectra are given 
in flux units 
of $10^{-15}$ erg \cm2\ s$^{-1}$ \AA$^{-1}$. The detection, 
measurement and identification of all absorption lines have been performed
interactively.
The uncertainty on observed equivalent widths is derived (in the same manner 
as Young et al. 1979) from the noise level measured in selected portions of 
the spectra which look free from any absorption line. For weak unresolved 
features, $\sigma(W_\mathrm{obs})$ lies in the range 0.06 -- 0.10~\AA\ over 
most of all the G190H and G270H ranges (0.07 -- 0.08~\AA\ being by far the 
most common 
values). Locally, $\sigma(W_\mathrm{obs})$ can be smaller (e.g.
on emission lines where values as low as $\sigma(W_\mathrm{obs}) =
0.02$~\AA\ are reached) or larger (e.g. when blending occurs).
The central wavelengths of the absorption lines are obtained by fitting a
Gaussian to the observed profile (after normalizing the adjacent continuum if 
necessary, e.g. for absorption lines located on an emission line). For 
the identification of features significant at the $4.5\sigma$ level, 
we first search for absorption from already known metal systems using 
the list of strong lines given by Bahcall et al. (1993; their 
Table 1). The presence of weaker transitions that might be expected on the basis
of the results obtained in this first step is then examined, especially from the
DLAS, for which we accept lines at a lower significance level. 
To this aim, we consider the extensive line list 
given by Verner et al. (1994). The identification is performed on the basis of 
criteria involving redshift agreement, line width as compared to the 
instrumental 
profile and relative strength, when several transitions from the same
species are detected. When two or more lines strongly overlap, wavelengths and 
equivalent widths are measured after deblending. This was performed using the 
context FIT/LYMAN developed by A. Fontana within MIDAS, the ESO data analysis 
software package; this routine was also used to fit line profiles and extract 
gaseous column 
densities for damped \lya\ lines or Lyman edges. Among the lines left 
unidentified, we have searched for the presence of metal doublets and, 
in the \lya\ forest, for lines from the Lyman series possibly 
associated with the strongest candidate \lya\ features.

This results in the line lists given in Tables~\ref{q0302list} to 
\ref{q3c286list}.  For EX~0302$-$223, only lines from metal systems are given in
Table~\ref{q0302list}. Although the S/N ratio is quite good, some ambiguities 
remain 
due to 1) the large number of absorption features expected from the DLAS 
and 2) the presence in EX~0302$-$223, PKS~0454+039 and PKS~1229$-$021
of  several other metal systems. When two or more transitions might contribute 
to a given feature, they are indicated in the Tables. Le Brun \& Bergeron 
(1997) have performed 
an identification of \lya\ absorbers in the field of 3C~286 and give for this
object a more extensive list of candidate \lya\ lines (down to the $3\sigma$
significance level).
\section{\label{indiv}Comments on individual objects}
Here, we briefly summarize spectroscopic data previously obtained on our targets
and indicate the main results provided by the new spectra.
\begin{table*}[ht!]
\def\baselinestretch{1.}
\caption{\label{q0454list} Absorption lines detected in the spectrum of
PKS~0454+039}
\scriptsize{
\begin{tabular}{rrrrrlr||rrrrrlr}
\hline\noalign{\smallskip}
N & $\lambda _\mathrm{obs}$ & $\Wo$ & $\Wr$ & $\lambda _\mathrm{r}$ 
& Ident. & $\za$ &
N & $\lambda _\mathrm{obs}$ & $\Wo$ & $\Wr$ & $\lambda _\mathrm{r}$ 
& Ident. & $\za$ \\
\hline\noalign{\smallskip}
1  & 1721.81 & 1.47 &      & 1215.67 & \lya         & 0.4163 &
56 & 2243.97 & 1.38 & 0.74 & 1206.50 & Si\,{\sc iii}& 0.8599 \\
   &         &      &      &  926.23 & Ly$\eta$     & 0.8589 &
57 & 2260.90 &18.70 &10.05 & 1215.67 & \lya         & 0.8598 \\ 
2  & 1730.82 & 2.30 & 1.24 &  930.65 & Ly$\zeta$    & 0.8598 &
58 & 2287.45 & 1.00 & 0.53 & 1215.67 & \lya         & 0.8816 \\ 
3  & 1743.86 & 2.03 & 1.09 &  937.80 & \lye         & 0.8595 &
59 & 2319.98 & 1.13 & 0.59 & 1215.67 & \lya         & 0.9084 \\ 
4  & 1766.09 & 2.70 & 0.94 &  949.74 & \lyd         & 0.8596 &
60 & 2332.12 & 1.50 & 0.78 & 1215.67 & \lya         & 0.9183 \\
5  & 1782.40 & 0.54 & 0.37 & 1215.67 & \lya         & 0.4662 &
61 & 2343.76 & 1.85 &      & 1260.42 & Si\,{\sc ii} & 0.8595$^\mathrm{f}$ \\
6  & 1794.58 & 3.30 & 1.53 &  832.93 & O\,{\sc iii} & 1.1545$^\mathrm{a}$ &
   &         &      &      & 2344.21 & Fe\,{\sc ii} &-0.0002 \\
7  & 1808.54 & 3.90 & 2.10 &  972.54 & \lyc         & 0.8596 &
62 & 2356.43 & 1.00 & 0.51 & 1215.67 & \lya         & 0.9384 \\ 
8  & 1816.54 & 2.51 & 1.35 &  977.02 & C\,{\sc iii} & 0.8593 &
63 & 2367.94 & 0.59 & 0.29 & 1144.94 & Fe\,{\sc ii} & 1.0682$^\mathrm{c}$\\ 
9  & 1826.47 & 0.60 & 0.30 &  923.15 & Ly$\theta$   & 0.9785 &
64 & 2375.15 & 1.07 & 1.07 & 2374.46 & Fe\,{\sc ii} & 0.0002$^\mathrm{c}$\\
10 & 1832.29 & 0.85 & 0.43 &  926.23 & Ly$\eta$     & 0.9782 &
65 & 2382.42 & 1.11 & 1.11 & 2382.77 & Fe\,{\sc ii} &-0.0001 \\
11 & 1838.77 & 0.86 & 0.46 &  988.77 & O\,{\sc i}   & 0.8597 &
66 & 2404.78 & 3.23 & 1.63 & 1215.67 & \lya         & 0.9781 \\
12 & 1841.06 & 1.40 &      &  989.87 & Si\,{\sc ii} & 0.8599$^\mathrm{b}$ &
67 & 2421.57 & 1.25 & 0.67 & 1302.17 & O\,{\sc i}   & 0.8596 \\ 
   &         &      &      &  930.75 & Ly$\zeta$    & 0.9780 &
68 & 2424.97 & 2.38 &      & 1215.67 & \lya         & 0.9948 \\
13 & 1855.19 & 1.57 & 0.79 &  937.80 & \lye         & 0.9782 &
   &         &      &      & 1304.37 & Si\,{\sc ii} & 0.8591 \\ 
14 & 1878.62 & 1.57 & 0.79 &  949.74 & \lyd         & 0.9780 &
69 & 2481.75 & 1.35 & 0.73 & 1334.53 & C\,{\sc ii}  & 0.8596 \\ 
15 & 1893.75 & 0.57 & 0.36 & 1215.67 & \lya         & 0.5577 &
70 & 2499.23 & 1.02 & 0.50 & 1215.67 & \lya         & 1.0558 \\ 
16 & 1907.49 & 3.70 & 1.99 & 1025.72 & \lyb         & 0.8597 &
71 & 2513.90 & 0.85 & 0.41 & 1215.67 & \lya         & 1.0679 \\
17 & 1918.78 & 0.80 & 0.51 & 1031.93 & O\,{\sc vi}  & 0.8594 &
72 & 2543.37 & 0.64 & 0.31 & 1215.67 & \lya         & 1.0922 \\ 
18 & 1923.85 & 1.28 & 0.65 &  972.54 & \lyc         & 0.9782 &
73 & 2552.30 & 0.83 & 0.40 & 1215.67 & \lya         & 1.0995 \\ 
19 & 1927.28 & 0.66 & 0.35 & 1036.34 & C\,{\sc ii}  & 0.8597 &
74 & 2561.57 & 0.66 & 0.32 & 1238.82 & N\,{\sc v}   & 1.0678 \\
20 & 1929.16 & 0.42 & 0.22 & 1037.62 & O\,{\sc vi}  & 0.8592 &
75 & 2563.75 & 0.44 & 0.20 & 1190.42 & Si\,{\sc ii} & 1.1537 \\
21 & 1932.64 & 1.00 & 0.63 & 1215.67 & \lya         & 0.5898 &
76 & 2569.85 & 0.58 &      & 1242.80 & N\,{\sc v}   & 1.0678 \\
22 & 1975.58 & 0.57 & 0.35 & 1215.67 & \lya         & 0.6251 &
   &         &      &      & 1193.29 & Si\,{\sc ii} & 1.1536 \\
23 & 1983.13 & 0.56 & 0.26 &  920.96 & Ly$\iota$    & 1.1533 &
77 & 2587.02 & 0.38 & 0.38 & 2586.65 & Fe\,{\sc ii} & 0.0001 \\ 
24 & 1988.08 & 1.24 & 0.58 &  923.15 & Ly$\theta$   & 1.1539 &
78 & 2591.92 & 0.92 & 0.49 & 1393.86 & Si\,{\sc iv} & 0.8595 \\ 
25 & 1995.00 & 1.12 & 0.52 &  926.23 & Ly$\eta$     & 1.1539 &
79 & 2599.00 & 1.42 &      & 2600.17 & Fe\,{\sc ii} &-0.0001 \\
26 & 1999.44 & 1.24 & 0.75 & 1215.67 & \lya         & 0.6447 &
   &         &      &      & 1206.50 & Si\,{\sc iii}& 1.1542 \\
27 & 2004.57 & 1.25 & 0.58 &  930.75 & Ly$\zeta$    & 1.1537 &
80 & 2607.68 & 0.63 &      & 1402.77 & Si\,{\sc iv} & 0.8591$^\mathrm{c}$\\
28 & 2010.54 & 0.80 & 0.39 &  972.54 & \lyc         & 1.0673 &
   &         &      &      & 1260.42 & Si\,{\sc ii} & 1.0689 \\
29 & 2015.87 & 1.80 & 0.97 & 1083.99 & N\,{\sc ii}  & 0.8597 &
81 & 2618.27 & 3.02 & 1.40 & 1215.67 & \lya         & 1.1538 \\ 
30 & 2019.73 & 1.84 & 1.05 &  937.80 & \lye         & 1.1537 &
82 & 2634.90 & 1.86 & 0.86 & 1215.67 & \lya         & 1.1674 \\ 
31 & 2028.78 & 2.26 & 1.12 & 1025.72 & \lyb         & 0.9779 &
83 & 2643.95 & 0.67 & 0.31 & 1215.67 & \lya         & 1.1749 \\ 
32 & 2040.63 & 0.47 & 0.28 & 1215.67 & \lya         & 0.6786 &
84 & 2652.83 & 1.23 & 0.56 & 1215.67 & \lya         & 1.1822$^\mathrm{c}$\\ 
33 & 2045.49 & 1.80 & 0.84 &  949.74 & \lyd         & 1.1537 &
85 & 2658.77 & 0.35 & 0.16 & 1215.67 & \lya         & 1.1871 \\ 
34 & 2062.79 & 0.32 & 0.19 & 1215.67 & \lya         & 0.6968 &
86 & 2667.41 & 0.61 & 0.28 & 1238.82 & N\,{\sc v}   & 1.1532 \\
35 & 2077.82 & 0.66 & 0.38 & 1215.67 & \lya         & 0.7092$^\mathrm{c}$ &
87 & 2678.63 & 1.19 &      & 1215.67 & \lya         & 1.2034$^\mathrm{c}$ \\
36 & 2086.61 & 0.45 & 0.24 & 1121.99 & Fe\,{\sc ii} & 0.8597$^\mathrm{c}$ &
   &         &      &      & 1242.80 & N\,{\sc v}   & 1.1553$^\mathrm{c}$ \\ 
37 & 2094.54 & 2.76 & 1.28 &  972.54 & \lyc         & 1.1537 &
88 & 2714.49 & 0.60 & 0.28 & 1260.42 & Si\,{\sc ii} & 1.1536 \\
38 & 2104.36 & 1.91 & 0.88 &  977.02 & C\,{\sc iii} & 1.1539 &
89 & 2724.17 & 1.20 & 0.54 & 1215.67 & \lya         & 1.2409 \\ 
39 & 2109.45 & 1.75 &      & 1134.66 & N\,{\sc i}   & 0.8591 &
90 & 2728.95 & 0.95 & 0.42 & 1215.67 & \lya         & 1.2448 \\ 
   &         &      &      & 972.54  & \lyc         & 1.1690 &
91 & 2735.56 & 0.32 & 0.14 & 1215.67 & \lya         & 1.2502 \\
40 & 2121.38 & 0.78 & 0.37 & 1025.72 & \lyb         & 1.0682 &
92 & 2741.05 & 0.53 & 0.23 & 1215.67 & \lya         & 1.2548 \\ 
41 & 2129.40 & 0.50 &      &  988.77 & O\,{\sc i}   & 1.1536 &
93 & 2764.14 & 0.60 & 0.26 & 1215.67 & \lya         & 1.2738 \\
   &         &      &      & 1144.94 & Fe\,{\sc ii} & 0.8598 &
94 & 2773.87 & 1.25 & 0.55 & 1215.67 & \lya         & 1.2818 \\
42 & 2131.87 & 0.62 &      &  989.80 & N\,{\sc iii} & 1.1538$^\mathrm{d}$&
95 & 2796.57 & 1.05 & 1.05 & 2796.35 & Mg\,{\sc ii} & 0.0001 \\ 
43 & 2133.94 & 0.36 & 0.17 & 1031.93 & O\,{\sc vi}  & 1.0679 &
96 & 2803.58 & 1.38 & 1.38 & 2803.53 & Mg\,{\sc ii} & 0.0000 \\ 
44 & 2135.50 & 0.51 & 0.29 & 1215.67 & \lya         & 0.7566 &
97 & 2839.12 & 1.49 & 0.80 & 1526.71 & Si\,{\sc ii} & 0.8596 \\ 
45 & 2144.87 & 1.30 & 0.74 & 1215.67 & \lya         & 0.7644$^\mathrm{e}$ &
98 & 2848.44 & 0.50 & 0.21 & 1215.67 & \lya         & 1.3431 \\
46 & 2180.18 & 0.20 & 0.11 & 1215.67 & \lya         & 0.7934 &
99 & 2852.75 & 0.87 & 0.87 & 2852.96 & Mg\,{\sc i}  &-0.0001 \\ 
47 & 2186.54 & 0.21 & 0.11 & 1215.67 & \lya         & 0.7986 &
100& 2873.88 & 1.01 & 0.47 & 1334.53 & C\,{\sc ii}  & 1.1535 \\
48 & 2196.62 & 0.70 & 0.39 & 1215.67 & \lya         & 0.8069 &
101& 2879.08 & 1.45 & 0.78 & 1548.20 & C\,{\sc iv}  & 0.8596 \\ 
49 & 2203.45 & 0.63 & 0.35 & 1215.67 & \lya         & 0.8125 &
102& 2883.81 & 1.01 & 0.54 & 1550.77 & C\,{\sc iv}  & 0.8596 \\
50 & 2208.96 & 2.97 & 1.38 & 1025.72 & \lyb         & 1.1536 &
103& 2991.49 & 0.90 & 0.48 & 1608.45 & Fe\,{\sc ii} & 0.8598 \\
51 & 2213.88 & 1.17 & 0.63 & 1190.42 & Si\,{\sc ii} & 0.8597 &
104& 2996.15 & 0.77 & 0.72 & 2796.35 & Mg\,{\sc ii} & 0.0715 \\
52 & 2219.17 & 1.07 & 0.57 & 1193.29 & Si\,{\sc ii} & 0.8597 &
105& 3001.56 & 0.85 & 0.39 & 1393.76 & Si\,{\sc iv} & 1.1536 \\
53 & 2222.93 & 1.66 &      & 1025.72 & \lyb         & 1.1672 &
106& 3003.81 & 0.71 & 0.66 & 2803.53 & Mg\,{\sc ii} & 0.0714 \\
   &         &      &      & 1031.93 & O\,{\sc vi}  & 1.1541 &
107& 3020.98 & 0.51 & 0.24 & 1402.77 & Si\,{\sc iv} & 1.1536 \\ 
54 & 2231.76 & 1.60 &      & 1199.55 & N\,{\sc i}   & 0.8605 &
108& 3107.24 & 1.17 & 0.63 & 1670.79 & Al\,{\sc ii} & 0.8597 \\ 
   &         &      &      & 1036.34 & C\,{\sc ii}  & 1.1535 &
109& 3201.83 & 0.57 & 0.28 & 1548.20 & C\,{\sc iv}  & 1.0681 \\
55 & 2234.84 & 0.38 &      & 1037.62 & O\,{\sc vi}  & 1.1538 &
110& 3206.98 & 0.45 & 0.22 & 1550.77 & C\,{\sc iv}  & 1.0680 \\
\noalign{\medskip}\hline
\end{tabular}
\smallskip\par
$^\mathrm{a}$ also O\,{\sc ii}834\\
$^\mathrm{b}$ also N\,{\sc iii}989\\
$^\mathrm{c}$ blend\\
$^\mathrm{d}$ also \siii 989\\
$^\mathrm{e}$ blend with \ovi\ at $z=1.0680$\\
$^\mathrm{f}$ also S\,{\sc ii}1259 \\
}
\def\baselinestretch{1.}
\end{table*}                                           
\subsection{EX~0302$-$223}
Metal-rich absorption systems have been mentioned for this object 
at $\za = 0.4196$ (\mgii: Bergeron, unpublished) and $\za = 1.0095$ (\feii\ 
and \mgii: Petitjean \& Bergeron 1990). 
The FOS spectrum (Fig.~\ref{spec0302}), clearly shows that the \lya\ line from the latter is damped
and reveals several strong features from \cii, \civ, \ni, \oi,
\siii, \siiii\ and \siiv. No less than four new metal systems are detected. 
In agreement with the prediction made in paper I, relatively strong lines 
from \feii\ and \mgii\ are seen at a redshift $\za = 0.1179$, similar to that 
of a bright spiral galaxy located $43.8 \h50$~kpc from the QSO line of sight. 
Second, a strong \civ\ doublet together with \lya\ 
is present at $\za = 0.9109$. In addition to \lya, \lyb\ and \lyc\, the FOS
spectrum shows weak features from \cii, \ciii, \siii\ and \siiii\ at $\za =
1.3284$; this system induces a Lyman edge near 2120~\AA\ which is clearly 
apparent in the IUE spectrum (Lanzetta et al.
1995). On the other hand, no convincing line 
is seen from the $\za = 0.4196$ \mgii\ system (the line at
2371~\AA\ could be \alii 1670 at $z=0.4191$, but this feature seems to be too 
strong given that we do not detect \feii 1608). Finally, a narrow high 
ionization
system with strong \ovi\ lines is detected at $\za = 1.4055 \simeq \ze$. 
The HST data do not confirm the $\za = 0.9690$ and 
0.9874 candidate damped \lya\ lines proposed by Lanzetta et al. (1995) (the 
former turns out to be \lyb\ at 1.3284).
\subsection{PKS 0454+039}
Before this study, two absorption systems were known in this object, at 
$\za= 0.8596$ and 1.1537 (Burbidge et al. 1977, Caulet 1989; Steidel \& Sargent 
1992). 
Steidel et al. (1995) have shown that the first one is a low metallicity 
(about 1/10 solar) DLAS, as indeed suggested by the strength of \mgii\
and \feii\ lines,
while in the second system, the \lya\ line is not damped. More than twenty 
metal lines are seen at 
$\za= 0.8596$ together with a Lyman edge near 1700~\AA\ (Fig.~\ref{spec0454}). 
The \siii 1304 line is definitely blended with another (\lya $-$only) line
since i) it appears too strong with respect to e.g. \siii1260 and ii) the match 
in redshift
is not satisfactory. In addition to several metal lines from \siii, \siiii, 
\siiv, \cii\ and \ciii, the $\za=1.1537$ system displays a beautiful set of 9 
lines from the Lyman series ending with a partial Lyman edge; the presence 
of \nv\ and \ovi\ lines in this system remains uncertain. 
One additional motivation for observing this QSO is the presence of 
two intervening galaxies detected by Steidel et al. (1993). 
The closest one is a dwarf star-forming galaxy at $\zg = 0.072$ from which we do
detect \mgii\ absorption. The second galaxy is a more luminous one at 
$z = 0.2011$ and $D = 118\h50$~kpc which could induce \civ\ absorption; 
no lines are seen from this galaxy but we detect a new
\civ\ doublet at $z=1.0682$.
\begin{figure*}
\centerline{\psfig{figure=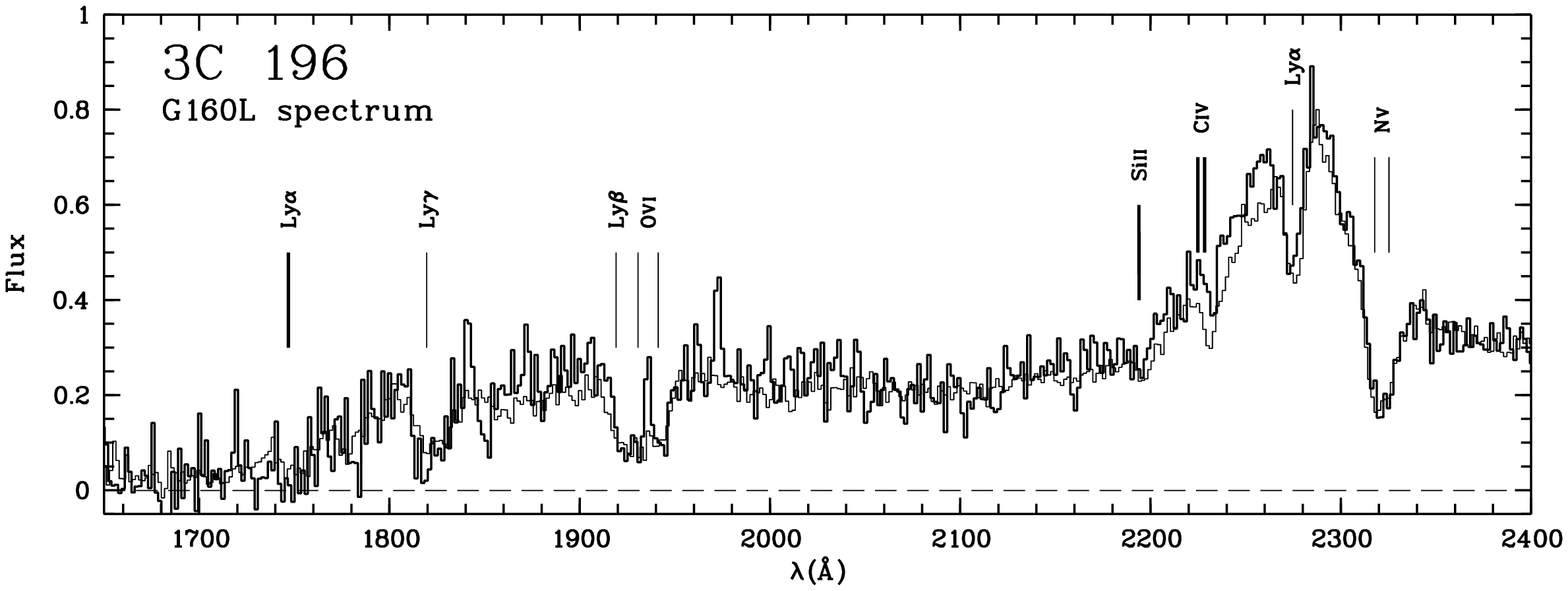,width=18cm,angle=-90,clip=true}}
\caption{\label{spec3c196} G160L spectra of 3C~196 obtained in 1994 as part of 
our program (thick line) and in 1992 by Cohen and collaborators (thin line) 
after correction of the wavelength scale (see text). The main absorption 
features from the systems at $\za = 0.437$ (DLAS) and $\za = 0.871$ are 
indicated (thick and thin tick marks respectively). Note the difference between 
the two spectra in the blue wing of the \lya\ emission line}
\end{figure*}
\subsection{3C~196}
The 21~cm absorption system at $\za = 0.4368$ is known to display very strong 
associated \feii\ and \mgii\ lines as well as \mnii\ and \caii\ (Foltz et al. 
1988; Boiss\'e \& Boulade 1990). An additional system is present at $\za 
\simeq \ze$. The G160L HST data presented by Cohen et al. (1996) show 
that, unfortunately, the latter produces a Lyman edge nearly coincident with the
\lya\ line at $\za = 0.437$, which renders the determination of $\nhi$ very difficult. 
An additional source of uncertainty in the analysis is related to the possible 
contribution of scattered light which makes the zero of the intensity scale ill 
defined. Our new (post-COSTAR) spectrum, although of lower integration time, 
is of interest in this regard. In order to perform a quantitative comparison, 
we retrieved from the HST archive the spectrum obtained in 1992 and analyzed
by Cohen et al. (1996).
Using the few narrow lines detected at either $\za = 0.000$, 0.437 or 0.871 
(Fig.~\ref{spec3c196}), we measure shifts of 14 and 6~\AA\ in the wavelength 
scales of the late and new spectrum respectively (these values are not accurate 
but the relative shift of 8~\AA\ is well constrained by the data). Once 
corrected, the two spectra appear in quite good agreement but a systematic 
difference is seen in the blue wing of the \lya\ emission line 
(Fig.~\ref{spec3c196}). If not instrumental, such an effect could be due to 
variable \lya\ absorption from gas ejected by the QSO; the time elapsed between 
the two observations is 2.05 yr and during that interval the amount or 
ionization degree of the gas may have changed (see e.g. Schartel et al. 1997 
for another example of variable absorption). An alternative possibility is 
intrinsic emission line variability.
Variable BAL-type absorption is supported by two facts: i) the gas responsible 
for the $\za = 0.871$ narrow absorption (to which the higher velocity gas is 
probably associated) is known to cover only partially the broad line region 
(Cohen et al. 1996) and 
ii) only a small variation of $\delta \nhi$ of about $2. \times 10^{15}$\cm2\ 
is needed to account for the strength of the effect (assuming optical 
thinness).
\subsection{Q~1209+107}
\begin{figure*}
\centerline{\psfig{figure=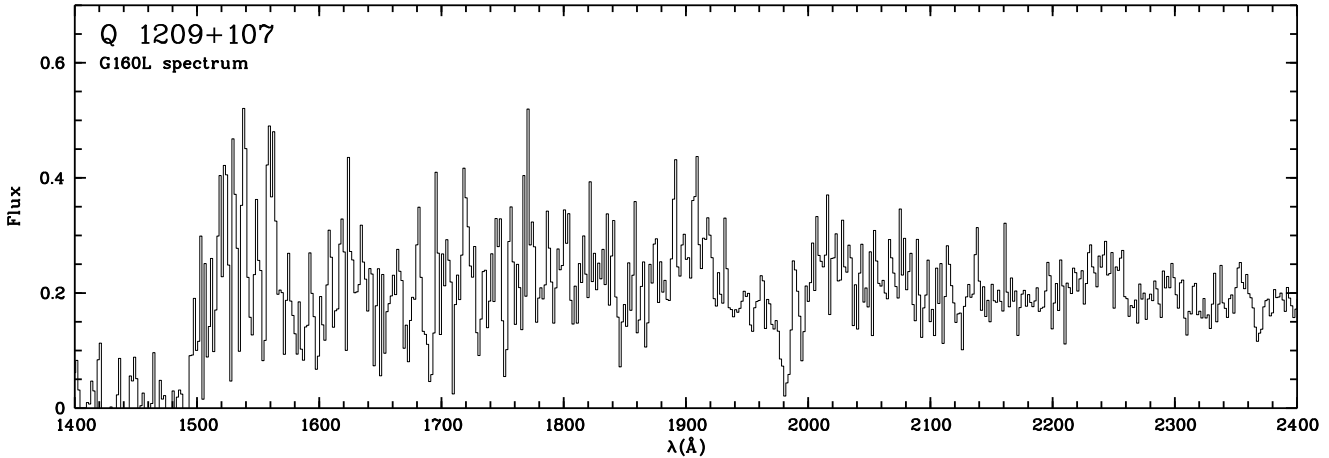,width=18cm,angle=-90,clip=true}}
\caption{\label{spec1209} G160L spectrum of Q1209+107}
\end{figure*}
The G160L FOS spectrum (Fig.~\ref{spec1209}), 
although of relatively poor S/N ratio, provides significant information on
the already known metal systems at $\za = 0.3930$, 0.6295 and 1.8434. First, 
these data confirm that the 0.6295 system is damped (see 
Sect.~\ref{q1209abond});
additional narrow absorptions (\lyb\ and \ni 1200) as well as a Lyman break 
are detected at this redshift. \lya\ at 0.3930 is also present and there are a 
few other possible features at 1.8434 (O\,{\sc iii}702 at 1995~\AA\ and
O\,{\sc iii}833 at 2369~\AA).  
\begin{figure*}
\centerline{\psfig{figure=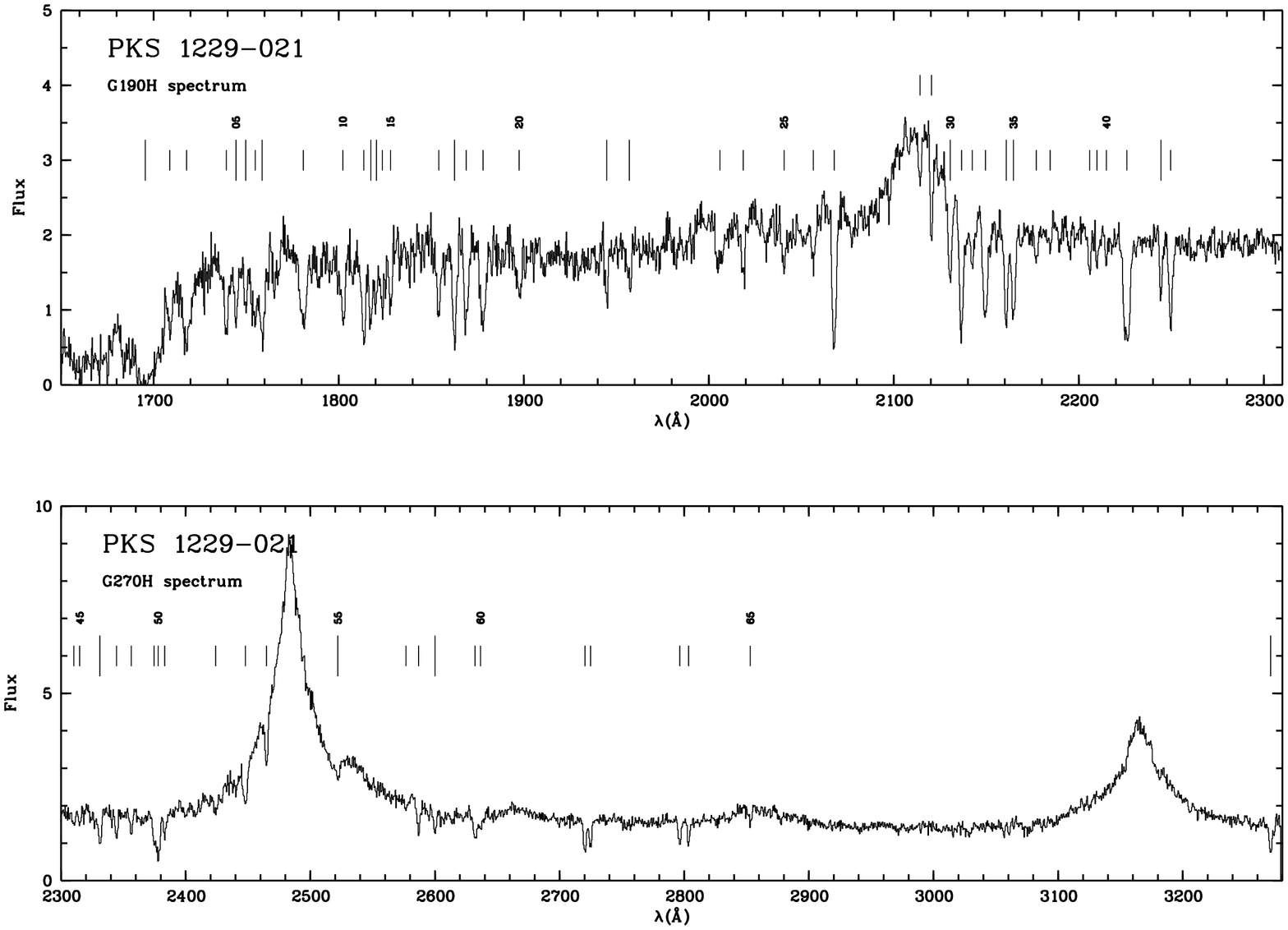,width=18cm,angle=-90,clip=true}}
\caption{\label{spec1229} Same as Fig.~\ref{spec0302} for PKS~1229$-$021 
(G190H and G270H spectra)}
\end{figure*}
\begin{table}
\def\baselinestretch{1.}
\caption{\label{q1229list} Absorption lines detected in the spectrum 
PKS~1229$-$021}
\smallskip
\scriptsize{
\begin{tabular}{rrrrrlr}
N & $\lambda _\mathrm{obs}$ & $\Wo$ & $\Wr$ & $\lambda _\mathrm{r}$
& Ident. & $\za$ \\
\hline\noalign{\smallskip}
  1 & 1695.56 & ??.? & 12.5 & 1215.67 & \lya         & 0.3948 \\
  2 & 1708.78 & 0.74 & 0.42 &  972.54 & \lyc         & 0.7570 \\
  3 & 1717.86 & 2.83 &      &  937.80 & \lye         & 0.8318$^\mathrm{a}$\\
    &         &      &      &  977.02 & \ciii        & 0.7583 \\
    &         &      &      &  972.54 & \lyc         & 0.7664 \\ 
  4 & 1739.35 & 1.18 & 0.64 &  949.74 & \lyd         & 0.8313 \\
  5 & 1744.53 & 0.77 &      & 1025.72 & \lyb         & 0.7008 \\
    &         &      &      & 1250.58 & S\,{\sc ii}  & 0.3950 \\
  6 & 1749.79 & 0.58 & 0.42 & 1253.81 & S\,{\sc ii}  & 0.3956 \\ 
  7 & 1754.95 & 1.25 & 0.87 & 1215.67 & \lya         & 0.4436$^\mathrm{a}$\\
  8 & 1758.70 & 1.88 & 1.35 & 1260.42 & Si\,{\sc ii} & 0.3953$^\mathrm{b}$\\
  9 & 1780.88 & 2.12 & 1.16 &  972.54 & \lyc         & 0.8312 \\
 10 & 1802.30 & 1.54 & 0.88 & 1025.72 & \lyb         & 0.7571 \\
 11 & 1813.61 & 1.76 &      & 1031.93 & O\,{\sc vi}  & 0.7575 \\
    &         &      &      & 1025.72 & \lyb         & 0.7681 \\ 
 12 & 1817.38 & 1.26 & 0.91 & 1302.17 & O\,{\sc i}   & 0.3957 \\
 13 & 1820.45 & 0.76 & 0.54 & 1304.37 & Si\,{\sc ii} & 0.3957 \\
 14 & 1823.60 & 1.11 & 0.63 & 1037.62 & O\,{\sc vi}  & 0.7575 \\
 15 & 1828.00 & 0.88 & 0.59 & 1215.67 & \lya         & 0.5037 \\
 16 & 1854.10 & 1.40 &      & 1854.72 & Al{\sc iii}  &-0.0003$^\mathrm{a}$\\
 17 & 1862.65 & 1.85 &      & 1334.53 & C\,{\sc ii}  & 0.3957$^\mathrm{c}$\\
    &         &      &      & 1862.79 & Al{\sc iii}  &-0.0001 \\
 18 & 1868.90 & 1.60 & 1.00 & 1215.67 & \lya         & 0.5373 \\
 19 & 1878.05 & 2.00 & 1.09 & 1025.72 & \lyb         & 0.8310 \\ 
 20 & 1897.52 & 0.98 & 0.53 & 1025.72 & \lyb         & 0.8499 \\
 21 & 1944.84 & 0.68 & 0.49 & 1393.76 & Si\,{\sc iv} & 0.3954  \\
 22 & 1957.12 & 0.65 & 0.47 & 1402.77 & Si\,{\sc iv} & 0.3952 \\
 23 & 2006.05 & 1.20 & 0.61 & 1025.72 & \lyb         & 0.9557$^\mathrm{a}$\\
 24 & 2018.65 & 0.83 & 0.50 & 1215.67 & \lya         & 0.6605 \\
 25 & 2040.68 & 0.44 & 0.26 & 1199.55 & \ni          & 0.7012$^\mathrm{d}$\\
 26 & 2056.46 & 0.52 & 0.31 & 1215.67 & \lya         & 0.6916 \\ 
 27 & 2067.78 & 2.47 & 1.45 & 1215.67 & \lya         & 0.7009 \\
 28 & 2114.14 & 0.38 & 0.22 & 1215.67 & \lya         & 0.7391 \\
 29 & 2120.39 & 0.71 & 0.40 & 1206.50 & Si\,{\sc iii}& 0.7575 \\
 30 & 2130.50 & 1.25 & 0.90 & 1526.71 & Si\,{\sc ii} & 0.3955 \\
 31 & 2136.60 & 1.94 & 1.11 & 1215.67 & \lya         & 0.7575 \\ 
 32 & 2142.37 & 0.70 & 0.40 & 1215.67 & \lya         & 0.7623 \\ 
 33 & 2149.52 & 1.90 & 1.07 & 1215.67 & \lya         & 0.7682 \\
 34 & 2160.74 & 1.42 & 1.02 & 1548.20 & C\,{\sc iv}  & 0.3956 \\ 
 35 & 2164.63 & 1.36 & 0.98 & 1550.77 & C\,{\sc iv}  & 0.3958 \\ 
 36 & 2177.00 & 0.33 & 0.19 & 1238.82 & N\,{\sc v}   & 0.7573 \\ 
 37 & 2184.46 & 0.21 & 0.12 & 1242.80 & N\,{\sc v}   & 0.7577 \\
 38 & 2205.78 & 0.44 & 0.24 & 1215.67 & \lya         & 0.8144 \\ 
 39 & 2209.79 & 0.33 & 0.18 & 1215.67 & \lya         & 0.8177 \\ 
 40 & 2214.81 & 0.22 & 0.13 & 1260.42 & Si\,{\sc ii} & 0.7572 \\
 41 & 2225.91 & 3.22 & 1.76 & 1215.67 & \lya         & 0.8310 \\ 
 42 & 2244.32 & 0.73 & 0.52 & 1608.45 & Fe\,{\sc ii} & 0.3953 \\ 
 43 & 2249.60 & 1.44 & 0.78 & 1215.67 & \lya         & 0.8505 \\
 44 & 2310.24 & 0.37 & 0.19 & 1215.67 & \lya         & 0.9004 \\
 45 & 2314.96 & 0.37 & 0.19 & 1215.67 & \lya         & 0.9043 \\
 46 & 2331.13 & 1.40 & 1.00 & 1670.79 & Al\,{\sc ii} & 0.3950 \\ 
 47 & 2344.50 & 0.86 & 0.86 & 2344.21 & Fe\,{\sc ii} & 0.0001 \\
 48 & 2356.31 & 0.72 & 0.37 & 1215.67 & \lya         & 0.9383 \\ 
 49 & 2374.63 & 0.95 & 0.95 & 2374.46 & Fe\,{\sc ii} & 0.0001 \\
 50 & 2377.92 & 2.50 & 1.28 & 1215.67 & \lya         & 0.9561 \\ 
 51 & 2383.22 & 0.93 & 0.93 & 2382.76 & Fe\,{\sc ii} & 0.0002 \\
 52 & 2424.03 & 0.52 & 0.26 & 1215.67 & \lya         & 0.9940 \\ 
 53 & 2447.97 & 1.18 & 0.67 & 1393.76 & Si\,{\sc iv} & 0.7564$^\mathrm{a}$ \\
 54 & 2464.89 & 0.76 & 0.43 & 1402.77 & Si\,{\sc iv} & 0.7572 \\
 55 & 2522.14 & 0.38 & 0.22 & 1808.01 & Si\,{\sc ii} & 0.3950 \\
 56 & 2576.68 & 0.24 & 0.24 & 2576.88 & Mn\,{\sc ii} &-0.0001 \\
 57 & 2586.91 & 0.95 &      & 2586.65 & Fe\,{\sc ii} & 0.0001 \\
    &         &      &      & 1854.72 & Al\,{\sc iii}& 0.3946 \\
 58 & 2600.10 & 0.77 &      & 2600.17 & Fe\,{\sc ii} &-0.0000 \\ 
    &         &      &      & 1862.79 & Al\,{\sc iii}& 0.3948 \\
 59 & 2632.20 & 1.17 & 0.69 & 1548.20 & C\,{\sc iv}  & 0.7002 \\
 60 & 2636.58 & 0.63 & 0.37 & 1550.77 & C\,{\sc iv}  & 0.7002 \\
 61 & 2720.33 & 1.73 & 0.99 & 1548.20 & C\,{\sc iv}  & 0.7571 \\ 
 62 & 2724.85 & 1.24 & 0.71 & 1550.77 & C\,{\sc iv}  & 0.7571 \\
 63 & 2796.40 & 1.00 & 1.00 & 2796.35 & Mg\,{\sc ii} & 0.0000 \\ 
 64 & 2803.49 & 0.76 & 0.76 & 2803.53 & Mg\,{\sc ii} & 0.0000 \\ 
 65 & 2853.02 & 0.62 & 0.62 & 2852.96 & Mg\,{\sc i}  & 0.0000 \\ 
 66 & 3270.54 & 1.30 & 0.93 & 2344.21 & Fe\,{\sc ii} & 0.3952 \\ 
\noalign{\medskip}\hline
\end{tabular}
\smallskip\par
$^\mathrm{a}$ Blend\\
$^\mathrm{b}$ also S\,{\sc ii}1259\\
$^\mathrm{c}$ also also C\,{\sc ii}*1335\\
$^\mathrm{d}$ also N\,{\sc i}1200
}
\def\baselinestretch{2.}
\end{table}
\subsection{PKS 1229$-$021}
This object has already been the subject of several detailed optical studies. 
In particular, the high resolution data published by Lanzetta \& Bowen (1992)
suggest a high metallicity for the intervening gas since strong \mnii\ 
lines are detected. We indeed find a large number of metal absorptions from 
the system at $\za = 0.3950$. In 1991, one of us observed this object with the 
IUE in order to detect 
the damped \lya\ line and determine the \hi\ column density; surprisingly, 
near the expected wavelength of this feature, a cut-off was seen 
(this IUE spectrum is shown in the catalog of Lanzetta et al. 
1993). Steidel et al. (1994a) detected a \mgii\ doublet at 0.7570 and proposed
that this system is responsible for the observed Lyman edge. Our data indicate 
that 
this \mgii\ doublet belongs to an extensive metal line system with 
absorptions from \siii, \siiii, \siiv, \civ,
\nv\ and \ovi. We detect an additional \civ\ system at 
$\za = 0.7003$. Furthermore, a strong \lya\ line is seen at $\za = 0.8310$ 
with several associated lines from the Lyman series; a careful examination 
of the optical data published by Steidel et al. (1994a)  
suggested the presence of weak associated \mgii\ lines near 5120~\AA. 
Measurements performed 
on the spectrum that C. Steidel kindly communicated to us confirm that shallow 
\mgii\
absorption is indeed present (the 2796~\AA\ line is seen at 
$\lambda_{obs}=5120.0$~\AA\ and with $W_\mathrm{obs}=0.30$~\AA). 
On the basis of our higher resolution and better S/N ratio UV data,
we find  that the partial cut-off near 1670~\AA\ is in fact due to this
$\za = 0.8310$ system. Fortunately, the damped \lya\ line at $\za = 0.3950$ 
can nevertheless be seen, superimposed onto the 
attenuated continuum (Fig.~\ref{spec1229}). 
\begin{table}
\def\baselinestretch{1.}
\caption{\label{q3c286list}Absorption lines detected in the spectrum of 3C~286}
\smallskip
\scriptsize{
\begin{tabular}{rrrrrlr}
N & $\lambda _\mathrm{obs}$ & $\Wo$ & $\Wr$ & $\lambda _\mathrm{r}$
& Ident. & $\za$ \\
\hline\noalign{\smallskip}
  1 & 1645.63 & 3.0  & 1.77 &  972.54 & \lyc          & 0.6921 \\
  2 & 1679.80 & 0.90 & 0.55 & 1025.72 & \lyb          & 0.6377 \\
  3 & 1711.37 & 0.82 & 0.58 & 1215.67 & \lya          & 0.4078 \\
  4 & 1735.96 & 7.86 & 4.65 & 1025.72 & \lyb          & 0.6924 \\
  5 & 1753.86 & 0.28 & 0.16 & 1036.34 & C\,{\sc ii}   & 0.6924 \\
  6 & 1781.06 & 0.80 & 0.46 & 1025.72 & \lyb          & 0.7364$^\mathrm{a}$\\
  7 & 1799.72 & 1.01 & 0.68 & 1215.67 & \lya          & 0.4804 \\
  8 & 1813.43 & 0.93 & 0.62 & 1215.67 & \lya          & 0.4917 \\
  9 & 1903.44 & 0.40 & 0.26 & 1215.67 & \lya          & 0.5658 \\
 10 & 1911.47 & 0.86 & 0.55 & 1215.67 & \lya          & 0.5724 \\
 11 & 1990.53 & 1.45 & 0.88 & 1215.67 & \lya          & 0.6374 \\
 12 & 2055.83 &51.9  & 30.7 & 1215.67 & \lya          & 0.6921 \\
 13 & 2111.73 & 1.03 & 0.59 & 1215.67 & \lya          & 0.7371 \\
 14 & 2133.00 & 0.79 & 0.47 & 1260.42 & Si\,{\sc ii}  & 0.6923$^\mathrm{b}$\\ 
 15 & 2143.86 & 0.30 & 0.17 & 1215.67 & \lya          & 0.7635 \\
 16 & 2149.72 & 0.83 & 0.47 & 1215.67 & \lya          & 0.7683 \\ 
 17 & 2152.64 & 0.34 & 0.19 & 1215.67 & \lya          & 0.7707 \\
 18 & 2179.61 & 0.28 & 0.16 & 1215.67 & \lya          & 0.7929 \\
 19 & 2183.11 & 0.52 & 0.29 & 1215.67 & \lya          & 0.7958 \\
 20 & 2203.60 & 0.28 & 0.16 & 1302.17 & O\,{\sc i}    & 0.6923 \\
 21 & 2207.59 & 0.24 & 0.14 & 1304.37 & Si\,{\sc ii}  & 0.6925 \\
 22 & 2224.00 & 0.24 & 0.14 & 1215.67 & \lya          & 0.8294 \\
 23 & 2583.78 & 0.29 & 0.15 & 1526.71 & Si\,{\sc ii}  & 0.6924 \\
 24 & 2587.32 & 0.27 & 0.27 & 2586.65 & Fe\,{\sc ii}  & 0.0003 \\
 25 & 2600.13 & 0.41 & 0.41 & 2600.17 & Fe\,{\sc ii}  &$-$0.0000 \\
 26 & 2620.53 & 0.28 & 0.17 & 1548.19 & C\,{\sc iv}   & 0.6926 \\
 27 & 2722.13 & 0.24 & 0.14 & 1608.45 & Fe\,{\sc ii}  & 0.6924 \\
 28 & 2796.39 & 0.98 & 0.98 & 2796.35 & Mg\,{\sc ii}  & 0.0000 \\ 
 29 & 2803.51 & 1.62 & 1.62 & 2803.53 & Mg\,{\sc ii}  &$-$0.0000 \\
 30 & 2852.99 & 0.24 & 0.24 & 2852.96 & Mg\,{\sc i}   & 0.0000 \\ 
\noalign{\medskip}\hline
\end{tabular}
\smallskip\par
$^\mathrm{a}$ blend\\
$^\mathrm{b}$ also S\,{\sc ii}1259
}
\def\baselinestretch{2.}
\end{table}
\subsection{3C 286}
\begin{figure*}
\centerline{\psfig{figure=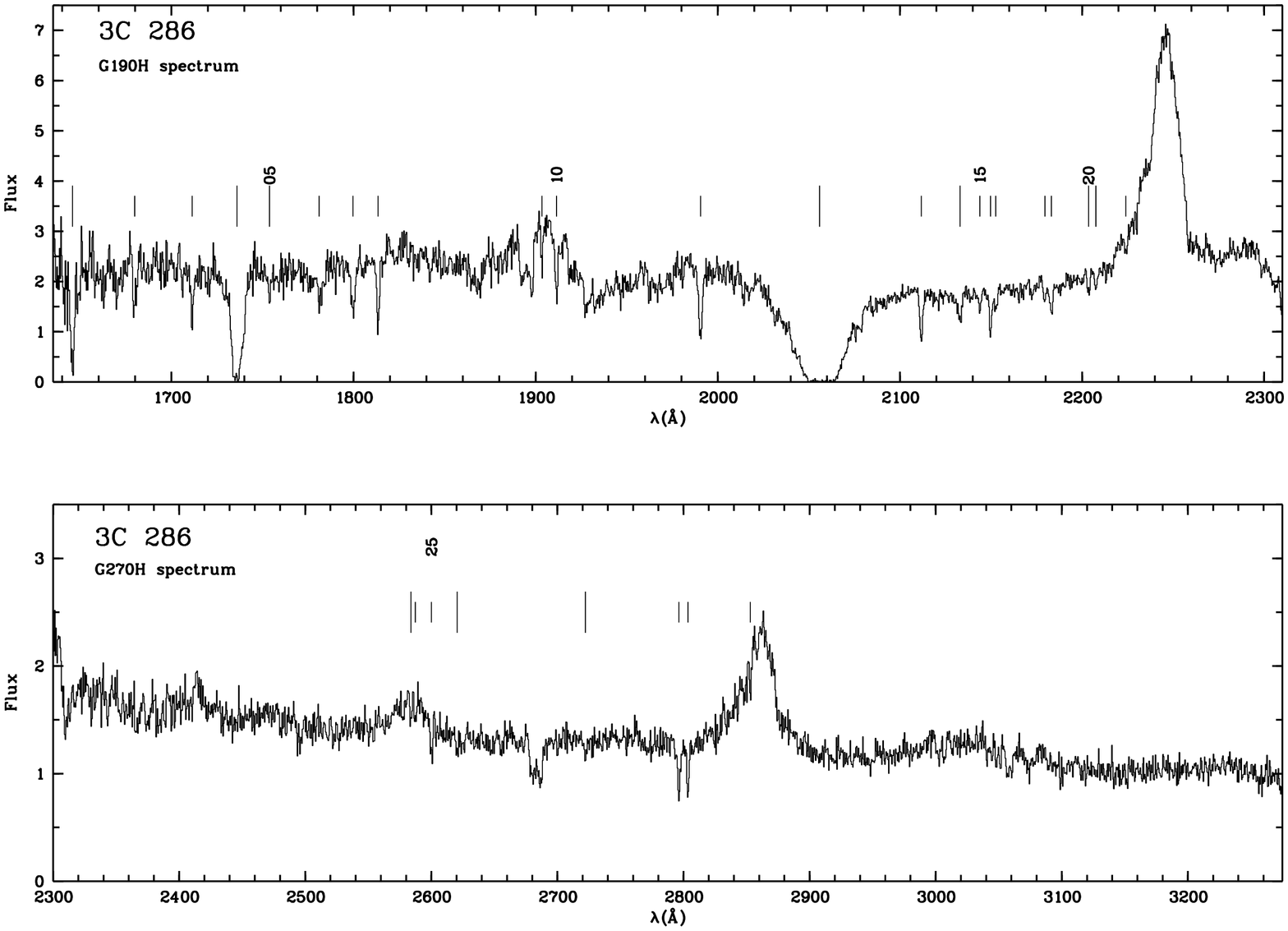,width=18cm,angle=-90,clip=true}}
\caption{\label{spec3c286} Same as Fig.~\ref{spec0302} for 3C~286 
(G190H and G270H spectra)}
\end{figure*}
We do not detect any additional narrow metal system in the new FOS spectrum 
(Fig.~\ref{spec3c286}), which makes
the line identification much easier for this QSO than for the other ones 
observed at high resolution. The profiles of the \ovi, \nv\ and to a lesser 
extent, of the \lya\ and \civ\ emission lines suggest the presence 
of broad absorption from highly ionized gas at $\za \simeq \ze$. In particular, 
two sharp edges are seen near 1925 and 2310~\AA\ \ which, when attributed to 
the \ovi\ and \nv\ doublets, correspond to about the same redshift, 
$\za = 0.865$,
that is an infall velocity of 2700\kms\ relative to the QSO. 
From the DLAS, we detect several new (mostly low-ionization) species
in addition to those - \feii, \mgii, \mgi, \znii, \crii\ and 
\caii\ - already seen by Meyer \& York (1992) and Cohen et al. (1994).
Given the strength of the \lya\ line, the \civ\ doublet is remarkably weak; the
second (1551~\AA) doublet line can barely be seen. \cii 1335 lies just
at the red end of the \lya\ emission line and its equivalent width strongly depends 
on the adopted shape for the adjacent continuum ($\Wo = 0.25$\AA\
in the normalized spectrum presented in Fig.~\ref{q3c286ci}).
Weak \ni 1200 and \siiii 1206 absorption can also be seen in the blue 
wing of the damped \lya\ line but 
their significance is difficult to assess quantitatively.
Doublets from \ovi, \nv\ and \siiv\ are not detected, with good upper limits.
\section{The properties of intermediate redshift damped \lya\ systems}
\subsection{\hi\ column density and metal abundances}
In the following, we discuss the determination of $\nhi$ and metal abundances
for each DLAS. Regarding metal lines, the type of data presented here are not 
appropriate to
decide whether they are optically thin or not. We then rely on the velocity 
distribution 
inferred from high resolution optical spectra (for species with a similar ionization 
level), when 
available. In order to get measurements or limits for several metals
from a given system, we use as much as possible already published optical 
spectra. A curve of
growth analysis is performed in cases where no optically thin line has been 
detected.
For consistency with
previous studies, we adopt the $f$ values given by Morton (1991) and follow the 
revisions proposed by Tripp et al. (1996). Abundances are expressed in terms of 
$[X/Y] = \log(X/Y) - \log(X/Y)_\odot$ where $(X/Y)_\odot$ represents 
Solar abundances taken from Anders \& Grevesse (1989).\\
In order to improve our ability to measure the abundance of minor elements, we 
adapt 
the stacking technique already used to search for specific features within a 
whole
class of absorption systems (e.g. \civ\ lines from \lya\ forest systems: Lu 
1991).
Some elements like \niii\ display several transitions in the same wavelength 
range
with comparable $f$ values and which are likely optically thin. It turns out 
that in the FOS
spectra presented in this paper, the resolution and S/N achieved are such that 
these
lines are occasionally seen individually but at a marginal significance level. 
We then extract a portion of the spectrum centered on the position where each 
such 
line from a given ion is expected and apply appropriate shifts so as to bring 
the various features
at the same wavelength (the shifts are computed a priori, from the absorption 
redshift
and rest wavelengths of the transitions). The individual spectra associated with 
each line 
are then averaged. In the optically thin limit, the column density can be 
obtained 
from the formula 
$$N = 1.13 \times 10^{20}{\sum_i W_i\over \sum_i f_i \lambda_i^2}~\mathrm{cm}^{-2},$$
where $f_i$, $W_i$ and $\lambda_i$ denotes the oscillator strength, 
rest equivalent width and wavelength (in ~\AA) of the transitions considered
(note that in the averaged spectrum, the "composite line" is characterized by an
equivalent width $\Wo = \sum_i {Wobs,i\over n}$ where $n$ is the number of 
lines). 
For undetected lines, all the quoted upper limits are $3\sigma$ 
values.
\subsubsection{EX~0302$-$223}
\begin{figure}
\centerline{\psfig{figure=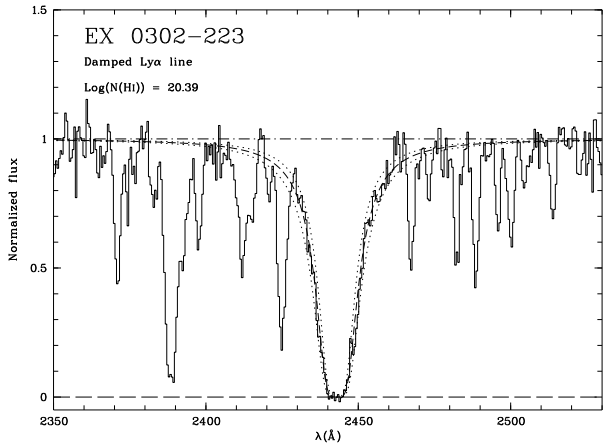,width=8cm,clip=t,angle=-90}}
\caption[]{\label{q0302dlas} Damped \lya\ line at $\za=1.0095$ in the spectrum 
of EX~0302$-$223. The best fit is plotted (dashed line; $\lnhi=20.39$) as well 
as the fits 
corresponding to a $3\sigma$ deviation (dotted lines; $\lnhi=20.27$ and 20.51)
} 
\end{figure}

The estimate of $\nhi$ in this DLAS is complicated by the presence of 
the QSO \ovi/\lyb\ emission line, just redward of the damped \lya\ line at 
$\za\ = 1.0095$. Thus, before fitting the latter, the
continuum has been defined in a way that leaves a
reasonably symmetrical \lya\ absorption line profile.
We get $\lnhi=20.39\pm 0.04$, where the quoted $1\sigma$ error is dominated 
by the uncertainty in the definition of the continuum level and by the 
possible presence of blended \lya\ forest lines (see Fig.~\ref{q0302dlas}). 
Our result is compatible with an independent estimate based on the 
same spectrum performed by Pettini \& Bowen (1997); these authors have measured 
the \znii\ and \crii\ column densities and derived [Zn/H] $= -0.5$ and [Cr/H] $= 
-0.9$.

In their high resolution spectra, Petitjean \& Bergeron (1990) have 
detected substructure in the \feii\ (2586 and 2600 \AA), \mgii\ and
\mgi\ lines and determine $\lnfeii = 14.29$. The HST spectrum provides
measurements for the 1144 and 1608~\AA\ \feii\ lines. Since in the high resolution data
\nfeii\ is dominated by one single component, a curve of growth analysis
is appropriate. The equivalent width of the 1144 and 1608~\AA\ lines 
appear to be well consistent 
with each other (both transitions have comparable $f$ values); 
when all four lines are considered, no unambiguous solution can be 
obtained and only a lower limit to the \feii\ column density can be inferred 
from the data, $\nfeii \ge 3.2 \times 10^{14}$ \cm2\ (i.e. $\lnfeii \ge 14.50$; a 
better fit is obtained with $\lnfeii$ in the range 14.6 -- 14.8 but the 1144 and 
1608~\AA\ line measurements are not accurate enough to provide a reliable upper 
limit).
We adopt the latter value in the following although it is larger 
than that given by Petitjean \& Bergeron (1990): their results
are based on the saturated 2586 and 2600~\AA\ \feii\ lines and are therefore subject to
large uncertainties (e.g. related to the unknown exact zero 
intensity level). We then infer [Fe/H] $\ge -1.40$. 
In the HST spectrum, \niii 1317 and 1370 are marginally
present ($2\sigma$ detection) with $\Wo \simeq 0.18$~\AA ; the corresponding
$3\sigma$ upper limit is [Ni/H] $< - 0.9$ (the stacking procedure is useless 
in this case since the portions of the spectra involved are crowded). 
Similarly, from the non detection
of \mnii\ by Petitjean \& Bergeron (1990), one gets [Mn/H]$< - 1.3$. 
Several lines from \siii\ are present but unfortunately, all have 
comparable and large oscillator strength values and cannot be used to 
reliably determine $\nsiii$.
\subsubsection{PKS 0454+039}
%
\begin{figure}
\centerline{\psfig{figure=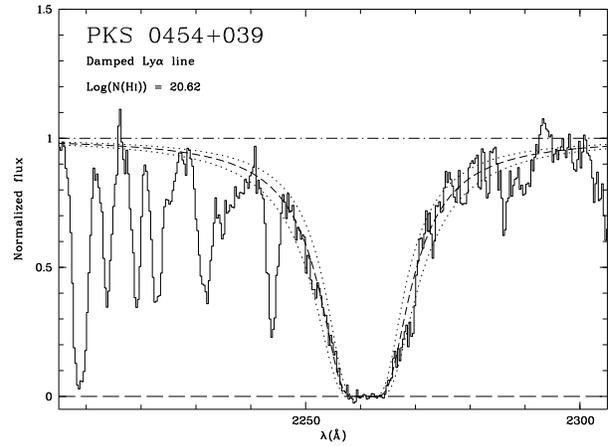,width=8cm,clip=t,angle=-90}}
\caption[]{\label{q0454dlas} Same as Fig.~\ref{q0302dlas} for the damped \lya\ 
line at $\za=0.8596$ in the spectrum of PKS~0454+039. The three fits shown 
correspond to 
$\lnhi=20.63$, 20.69 (best fit) and 20.75
} 
\end{figure}
When fitting the damped \lya\ profile (Fig.~\ref{q0454dlas}), we have excluded 
a narrow feature
near 2268~\AA\ (presumably a \lya $-$only line) which induces some asymmetry.
We then get $\lnhi=20.69 \pm 0.02$, a value slightly smaller than 
that obtained by Steidel et al. (1995), $\lnhi =20.75 \pm 0.03$.
The difference is likely to arise from our exclusion of the 2268~\AA\ feature;
the latter tends to broaden the profile and is less visible on the lower 
S/N spectrum analyzed by Steidel et al. (1995). 
These values are compatible given the 
formal errors quoted above (which, moreover, do not include the uncertainty in 
positioning the continuum).
Steidel et al. (1995) observed several weak transitions from \feii, \znii\ 
and \crii\ from which they determine the abundance of these three elements. 
Lu et al. (1996) present high resolution
and high S/N data on various lines from this system. 
Several components are detected, spread over
160\kms; the major one has a width (FWHM) of about 30\kms. Lu et al. (1996)
confirm that the 2249 and 2260~\AA\ \feii\ lines used by Steidel et al. (1995)
are optically thin.

\begin{figure}
\centerline{\psfig{figure=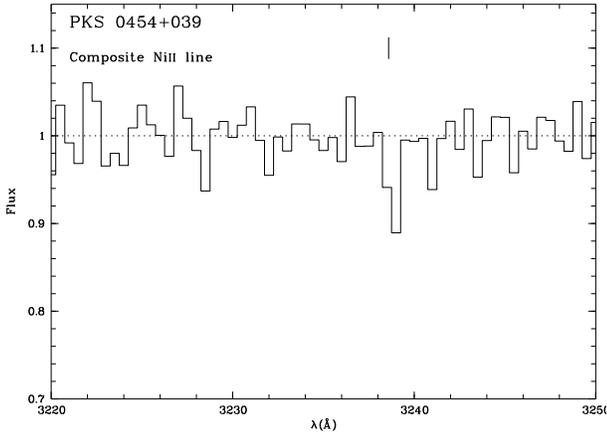,width=8cm,clip=t,angle=-90}}
\caption[]{\label{q0454ni} Composite \niii\ (1710, 1742 and 1752~\AA) line at 
$\za=0.8596$ in PKS~0454+039. Three portions of the original spectrum have been 
combined; the one including \niii 1710 and 1752 have been shifted by 59.5 
and $-19.5$~\AA\ before averaging. The tick mark indicates the expected 
position of the composite line} 
\end{figure}

Lines from \niii\ at 1710, 1742 and 1752~\AA\ are found to be 
marginally present in our spectrum and since they lie in reasonably clean
parts of the spectrum, we use the stacking technique discussed above. 
As can be seen in Fig.~\ref{q0454ni}, a $3\sigma$ feature is 
present at 3146.0~\AA, the expected wavelength of \niii1742 at 
$\za =$ 0.8596 with $\Wr =$ 0.18 $\pm$ 0.06~\AA.
This leads to a column density, $\lnniii = 13.36$, and a
relative abundance [Ni/H] $= -1.5$. 
The same procedure cannot be used in the more crowded region where 
\niii 1317 and 1370 are expected;
nevertheless, we check that these two features are individually marginally 
present 
(at about a $2\sigma$ level) with a strength compatible with the previous estimate.
For \mnii, Lu et al. (1996) get [Mn/H] $= -$1.36.
\subsubsection{\label{q3c196abond}3C 196}
As mentioned above, the damped \lya\ line at 0.437 
coincides with a Lyman edge and, in such a circumstance,
the low resolution data available poorly constrains $\nhi$. 
To better assess how far the value already inferred by Cohen et al. (1996)
depends on assumptions 
underlying the fitting procedure, we obtain an independent estimate 
of $\nhi$ based on our new spectrum. In the latter, the damped \lya\ line 
goes down to zero at its center and there is no need to correct for 
the presence of scattered light. Our approach has been to introduce the 
minimum number of parameters. Three at least are required: the $N$ and $b$ 
values 
for \hi\ at $z=0.871$ and $\nhi$ at $z=0.437$ (given the large $\nhi$ expected, 
the \lya\ profile is not expected 
to depend on the velocity distribution). We do not attempt to fit the \lya\ 
line at $z=0.871$ since Cohen et al. have shown that this requires the 
introduction of
an additional parameter - the fraction $f$ of the broad line region covered by 
the absorbing gas (which has to be less than 1); the 
relative contribution of emission lines shortward of $\lambda \simeq 2000$~\AA\ 
is expected to be small, so $f$ is no longer relevant. The strength of \lyb, 
\lyc, \lyd\ and of the Lyman discontinuity can be used to constrain $\nhi$ and 
$b$ at 0.871. We compute synthetic spectra for various ($\nhi$, $b$) values, 
degrade them to the resolution of the G160L spectra and compare them to the 
data. 

\begin{figure}
\centerline{\psfig{figure=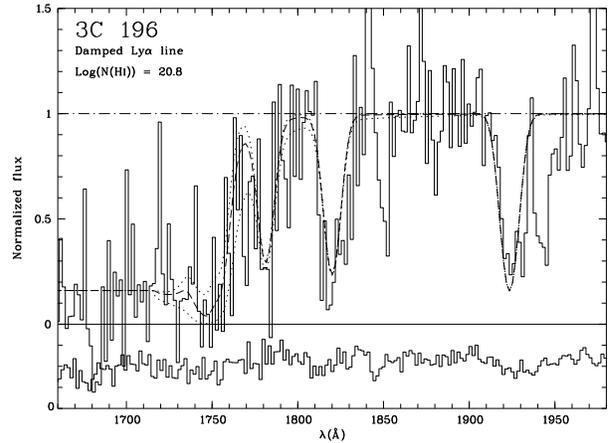,width=8cm,clip=t,angle=-90}}
\caption[]{\label{q3c196dlas} Same as Fig.~\ref{q0302dlas} for the damped \lya\ 
line at $\za=0.4368$ in the new (1994) spectrum of 3C 196. The three fits shown
correspond to $\lnhi =20.2$, 20.8 (best fit) and 21.4. The $1\sigma$ error is 
displayed and has been shifted downwards for clarity (bottom panel)
}
\end{figure}

We find that $\lnhi = 17.5$ and $b = 400$\kms\ roughly account for the 
strength of \lyb, \lyc, \lyd\ and for the possibly 
non-zero flux seen shortward of 
1700~\AA\ (note that the Cohen et al.'s spectrum also suggests a similar 
non-zero flux level, although the poor S/N ratio does not allow to be conclusive
on this point). In attempting to fit the observed spectrum, we find that 
the match is not quite so good for the wavelengths of \lyb\ and \lyd; since this 
is not the case for the Cohen et al.'s spectrum, we believe that this problem 
arises from
a lower S/N ratio or from distortions in the wavelength calibration and, 
to improve the fit, we allow slight wavelength shifts for these features.
This solution is certainly not unique (higher $\nhi$ and lower $b$ are also 
acceptable given the uncertainty in the flux level shortward of 1700~\AA; one 
must also
keep in mind that a single Gaussian may be a crude approximation of the real 
velocity
distribution); however, this is not critical since any choice within the 
acceptable range of values gives about the same shape for the 
edge when seen at our resolution. We then compute profiles for the \lya\ line
at 0.437, multiply these by the synthetic Lyman edge profile and compare the 
result 
to the data. We thus estimate $\lnhi ( \za = 0.437) = 20.8 \pm 0.2$ 
(the corresponding fits are displayed in Fig.~\ref{q3c196dlas}). 

This is notably larger than the value of 20.2 derived by Cohen et al. (1996). 
As we understand it, the 
difference comes from two reasons. Firstly, the two spectra show departures 
which, although relatively small, have large effects on the results. 
In the earlier spectrum, the damped \lya\ line is less deep and an intensity 
peak is present just shortward of it (near 1740~\AA) while this is much less 
clear in the latest data (this may be due to different line spread functions: a 
detailed comparison of the two spectra indicates that indeed, the latest has a 
significantly higher resolution). Secondly, in both spectra, the continuum 
is seen to fall off just shortward of 
1805~\AA, which cannot be due to the Lyman edge (the latter depresses the 
continuum only at $\lambda < 1765$ \AA). In their (low $\nhi$) solution, 
Cohen et 
al. gives a large weight to metal lines at $z=0.871$ (S\,{\sc vi} 933-944) and 
$z=0.437$ (\nv\ 1238-1242) which
induce the strong extra absorption required around 1766 and 1800~\AA. In our 
solution, this is naturally produced by the red wing of the 
damped \lya\ line itself. We find the large 
$\nhi$ solution more realistic because
we doubt S\,{\sc vi} at 0.871 and \nv\ at 0.437
can be as strong as required (note e.g. that the \siiv\ doublet at 0.437 is not 
detected). Further, it appears unlikely that the 
strength of these lines be precisely such that their cumulative effect produces 
the observed coherent fall off. However, although we favor a value above 20.5, 
we admit that the $\nhi$ value cannot be 
unambiguously determined with the present spectra; only higher resolution data
could allow to better model the Lyman edge at 0.871, 
assess the role of metal and \lya\ forest 
lines and determine the true profile of the damped \lya\ line around its core 
which really constrains $\nhi$.
 
Regarding metals, Foltz et al. (1988) have detected in the optical 
\mgii\ and \feii\ lines but the degree of saturation of the latter is 
such that they are useless for abundance estimates. One can nevertheless
get an upper limit on the \feii\ abundance from the non-detection of 
\feii 2367. With $f = 1.6 \times 10^{-4}$ and a 3$\sigma$ upper limit on $\Wr$ of
0.33~\AA, we get [Fe/H] $< 0.3$.
The weak features from \mnii\ and \caii\ are well resolved in the Foltz et al.'s
spectrum (line widths exceed 100\kms) and are therefore likely to be optically 
thin.
In this limit, we get $\lnmnii =  13.47$
from the intermediate strength 2595~\AA\ transition, thus [Mn/H] $= - 0.86$
(the two other \mnii\ transitions give consistent results; we also checked that 
the measurements performed by Aldcroft et al. (1994), 
although less accurate, are in acceptable agreement with those of Foltz et al. 1988). 
Similarly, from the \caii\ K line, we get $\lncaii = 12.78$. 
\subsubsection{\label{q1209abond}Q 1209+107}
%
\begin{figure}
\centerline{\psfig{figure=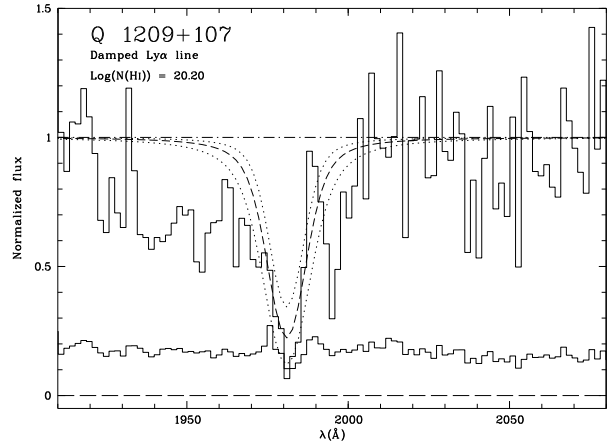,width=8cm,clip=t,angle=-90}}
\caption[]{\label{q1209dlas} Same as Fig.~\ref{q0302dlas} for the damped \lya\ 
line at $\za=0.6295$ in the spectrum of Q~1209+107. Fits are shown for 
$\lnhi =$ 19.9, 20.2 (best fit) and 20.5. The $1\sigma$ error is also given
} 
\end{figure}
Despite the low resolution and S/N, the spectrum constrains well the \hi\ 
column density at $\za = 0.6295$ and an acceptable fit to the \lya\ profile is 
obtained for $\lnhi = 20.2 \pm 0.1$ (Fig.~\ref{q1209dlas}). 
Acceptable fits can also be obtained by simultaneously decreasing \nhi\
and increasing b; however, such solutions are ruled out by the profile of 
the Lyman edge (Boiss\'e et al. 1998).
To our knowledge, 
the only metal lines from which 
an abundance can be derived for this system are those of \feii\ 
(Young et al. 1982). From a curve of growth analysis applied to five 
transitions, we get $\lnfeii = 14.82 \pm 0.08$, thus [Fe/H] $= - 0.9$.
Although no high resolution spectrum is available for Q~1209+107, we believe
that this determination is approximately correct because the line strengths
clearly indicate that \feii2374 lies close to the linear part of the curve of
growth (assuming this line to be thin yields the strict lower limit 
$\lnfeii > 14.70$). Young et al. (1982) do not detect the \mnii\ triplet;
we have used their $\sigma(W_\mathrm{obs})$ values to derive the limit
$\lnmnii \le 13.41$.
\subsubsection{\label{q1229abond}PKS 1229$-$021}
%
\begin{figure}
\centerline{\psfig{figure=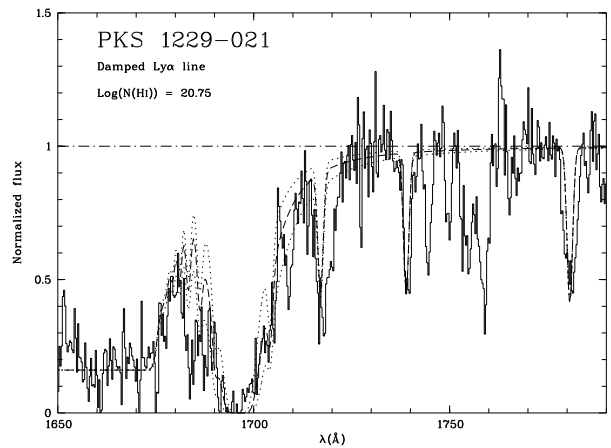,width=8cm,clip=t,angle=-90}}
\caption[]{\label{q1229dlas} Same as Fig.~\ref{q0302dlas} for the damped \lya\ 
line at $\zd=0.3950$ in the spectrum of Q~1229$-$021. Fits are shown for 
$\lnhi(\za=0.3950)=$ 20.54, 20.75 (best fit) and 20.96 
}
\end{figure}
As in 3C~196, the determination of $\nhi$ is complicated by the presence 
of an abrupt decrease of the continuum flux near the position of the 
expected damped \lya\ line due to a Lyman edge from the $\za = 0.831$ system.
However, the situation is more favorable than for 3C~196 because the spectral 
resolution is higher. We proceed as above 
and compute the profile for both the Lyman series/edge at 0.831
and the damped \lya\ at 0.3950. After assigning a zero intensity level 
to the core of the damped \lya\ (which requires a 7\% correction;
this is the only case for which the offset is negative), it is apparent
that the Lyman edge is not completely opaque. On the normalized spectrum, the 
level attained shortward of 1670~\AA\ is 0.17 from which we derive 
$\lnhi = 17.45$. We then attempt to reproduce \lya, \lyb, ... lines 
from this system by varying 
$b$: a single component does not provide a good fit to the data, the
observed lines being slightly too broad for their depth (the fit for 
$b = 34$~\kms\ is shown in Fig.~\ref{q1229dlas}). However, given the 
small velocity range involved, we have not attempted multi-component fits
because this would not affect the edge profile. 
In the HST spectrum, the fall off of the 
intensity begins at $\lambda \simeq 1730$~\AA : this is too large to be 
assigned to the $\za = 0.831$ edge but rather corresponds to the red wing of the 
damped \lya. In fact, at our resolution, the red half of the damped \lya\ line
is nearly unaffected by the Lyman edge which is favorable for the 
determination of $\nhi$. After successive trials, we get $\lnhi = 20.75 \pm 
0.07$.

PKS~1229$-$021 has been observed at high spectral resolution by Lanzetta \& 
Bowen (1992).
The velocity distribution appears complex and includes narrow components
spanning over 200\kms. In the FOS spectra, 
several lines that could be used for 
metal abundance measurements are expected.  From \siii, only \siii 1808 and 
\siii 1526 are of interest, other transitions being heavily blended. Assuming 
\siii 1808 to be optically thin, we get $\lnsiii = 15.54$, which is to be 
considered as a lower limit. Including \siii 1808 and \siii 1526 in a curve 
of growth analysis suggests that the former line is nearly thin and 
leads to $b = 46$\kms, a value roughly consistent with the profile observed for 
unsaturated lines by Lanzetta \& Bowen (1992) and $\lnsiii = 15.62$. However, 
since this
estimate may be affected by the presence of saturated narrow components in the 
\siii1526
line, we adopt for [Si/H] the thin limit, [Si/H] $> -$0.76.
\begin{figure}
\centerline{\psfig{figure=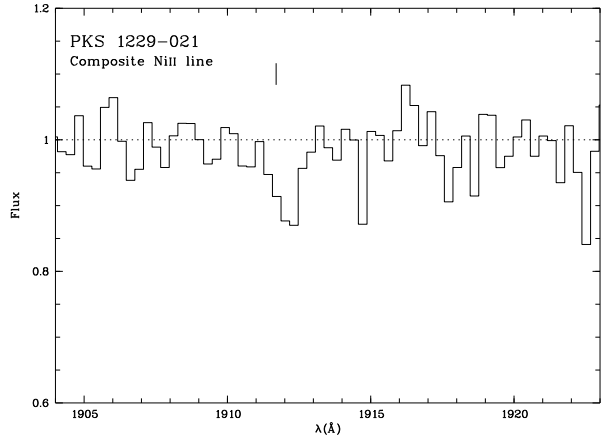,width=8cm,clip=t,angle=-90}}
\caption[]{\label{q1229ni} Same as Fig.~\ref{q0454ni} for the 
composite \niii\ (1317, 1370 and 1454~\AA) line at $\za=0.3950$ in the spectrum
of PKS~1229$-$021. The portions of the original spectrum including \niii1317 
and 1454 have been shifted by 74 and $-118$~\AA\ before averaging
}
\end{figure}

\begin{figure}
\centerline{\psfig{figure=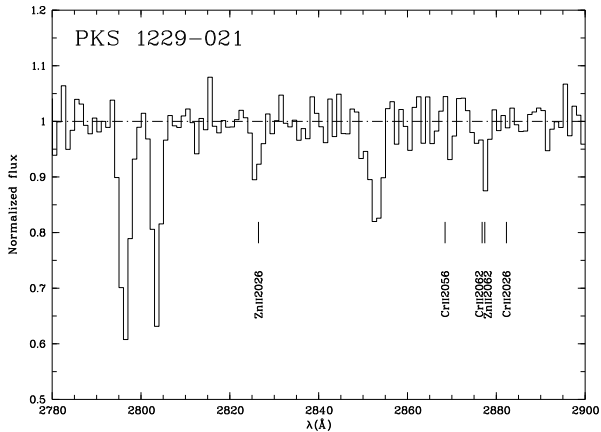,width=8cm,clip=t,angle=-90}}
\caption[]{\label{q1229znii} Portion of the PKS~1229$-$021 G270H spectrum
(binned to 1~\AA) comprising the \znii\ and \crii\ lines expected from the 
DLAS. The three strong lines near 2800 and 2850~\AA\ are from Galactic 
\mgii\ and \mgi
}
\end{figure}

Lanzetta \& Bowen (1992) observed the two \feii2586 and \feii2600 
lines. Since the latter are strongly saturated, their estimate of $\nfeii$  
heavily depends on the assumed velocity distribution (sum of discrete Gaussian 
components) and within this assumption, on the number of subcomponents introduced 
(in such a case, the approach developed by Levshakov \& Kegel 1997
to infer column densities may be interesting to consider). Indeed, their $\nfeii$ 
would lead to an unrealistically large Fe relative abundance and is in 
contradiction with
the non-detection in our spectrum of \feii2249 and 2260. By stacking the 
(assumed optically thin) two latter features, we get the upper 
limit $\lnfeii \le 14.94$. The corresponding relative abundance is [Fe/H] 
$\le -1.32$. We also stacked the \niii\ 1317, 1370 and 1454 \AA\ lines (the 
\niii\ lines above 1700~\AA\ fall near strong features and cannot be used) and 
the 
\crii\ 2056 and 2066 \AA\ lines. \niii\ is clearly detected (Fig.~\ref{q1229ni})
with $\left<W_\mathrm{i}\right> = 0.12$~\AA , which corresponds to 
$\lnniii = 13.70$ and [Ni/H] $= -1.3$. On the other hand, \crii\ is not present;
we get a 3$\sigma$ upper limit $\Wr \le 0.12$~\AA\ for the composite line 
which corresponds to $\lncrii \le 13.61$ or [Cr/H] $\le - 0.82$. 
Shallow features are seen at the expected position of \znii2026 and 2062 with
$\Wo = 0.21 \pm$ 0.06 \AA\ and $\Wo = 0.24 \pm$ 0.09 \AA\ respectively 
(Fig.~\ref{q1229znii}). Using 
the first measurement which is both more accurate and uncontaminated by 
absorption from other species (\mgi2026 is not expected to contribute 
significantly contrary to \crii2062), we get $\lnznii = 12.93 \pm 0.12$ 
and [Zn/H] $= -0.47$. 
From unsaturated \mnii\ lines, Lanzetta \& Bowen (1992) derive 
$\lnmnii = 13.45$ (sum of all subcomponents) which corresponds to [Mn/H] 
$= -0.83$ (since the three lines used are nearly thin over most of the profile, 
this
estimate is not subject to the large uncertainties previously mentioned for 
\feii; 
the thin limit yields $\lnmnii >  13.1$, 13.0 and 13.3 for \mnii2576, 2594 and 
2606
respectively). \caii\ H and K lines have been detected at 2~\AA\ resolution 
by Steidel et al. (1994a). These lines are also seen in an unpublished higher 
resolution spectrum (0.35~\AA\ FWHM) that P. Petitjean kindly made available
to us, with two components at $\za = 0.39497$ and 0.39516. Equivalent width 
values from these two spectra are in good agreement, and from the 
average $\Wo$ of \caii 3934 (0.27~\AA), we get $\lncaii = 12.34$.

\subsubsection{\label{q3c286abond}3C 286}
\begin{figure}
\centerline{\psfig{figure=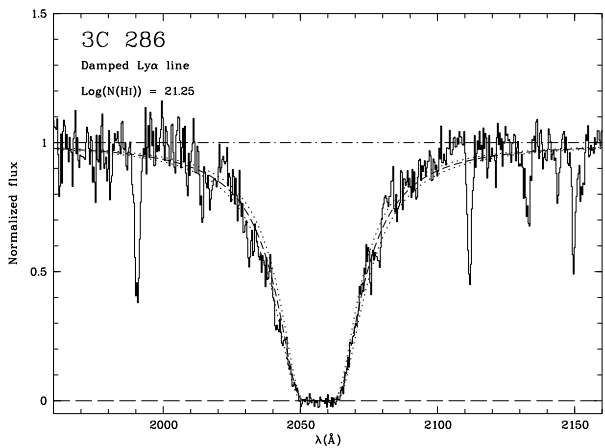,width=8cm,clip=t,angle=-90}}
\caption[]{\label{q3c286dlas} Same as Fig.~\ref{q0302dlas} for the damped \lya\ 
line at $\za=0.6922$ in the spectrum of 3C~286. The three fits shown correspond 
to 
$\lnhi=21.19$, 21.25 (best fit) and 21.31.
}
\end{figure}

The normalization of the spectrum near the damped \lya\ line
is uncertain due to the presence of adjacent emission lines. Therefore, when 
fitting
the profile, we give much weight to the core of the line and get $\lnhi = 21.25 
\pm 0.02$
which is in good agreement with the value $\lnhi = 21.29$ derived by Cohen et 
al.
(1994) from G160L data (Fig~\ref{q3c286dlas}). 
The DLAS in 3C~286 might seem to be a good case for abundance determinations 
since the velocity distribution comprises one single component with $b = 
6.5$\kms, 
a value which is consistent with both 21~cm and \feii\ data 
(see Meyer \& York 1992). However, such a low $b$ value implies high line 
opacities, even with abundances as low as 1/100 Solar. As a result, many of 
the new lines detected here all lie well beyond the linear part of the curve of 
growth 
despite their weakness. Since we have some a priori information on the velocity 
distribution we nevertheless attempt to derive \nsiii\ using the few transitions
for which reliable measurements could be made (\siii1260, 1304 and to a lesser 
extent,
\siii1526). A single Gaussian component with  $b = 6.5$\kms\ is clearly 
inconsistent 
with the data. \siii 1260 and 1304 could be accounted for if $b = 35$\kms\ and 
$\lnsiii = 14.20$ but such a large difference between $b(\mbox{\siii})$ and 
$b(\mbox{\feii})$ appears 
unlikely. Further, the corresponding relative Si abundance would be extremely 
low 
($\simeq -2.5$). It seems also that for any velocity distribution, 
\siii 1190 (which is barely seen) and especially 
\siii 1193 (undetected) should be notably stronger than observed, as compared to 
\siii 1260 or 1304; we may therefore suspect that one of the latter is affected 
by blending with a \lya $-$only feature and conclude that the present data do not 
allow to estimate \nsiii\ properly. S\,{\sc ii}1259 is clearly present on the 
blue wing
of \siii 1260 but higher resolution data would be needed to extract 
$N(\mbox{S\,{\sc ii}})$.

Constraints on $\nmnii$ can be derived from the absence of \mnii 2576 in the 
spectrum
obtained by Cohen et al. (1994). With $\sigma(\Wo) = 0.035~\AA$, one gets 
$\lnmnii \le 12.48$ in the thin (i.e. large b) limit; adopting $b=6.5$\kms\ 
instead 
yields a $3\sigma$ limit of 12.59 indicating that saturation effects
might be not negligible in this case. 
We therefore adopt the latter value which implies [Mn/H] $\le -2.1$. 
Regarding \feii, our measurement of \feii 1608 appears fully consistent with the 
curve of growth analysis given by Meyer \& York (1992) who derive $\lnfeii = 
14.95$.
The latter authors also give the abundance of \caii, $\lncaii = 12.48$.
Finally, the tighter constraint that we can get on $\nniii$ comes from our 
non-detection 
of \niii 1317, which is expected on the blue side of the \lya\ QSO emission line 
where
the noise level is low. The $3\sigma$ limit on $\Wr$ is 0.080~\AA; the thin 
limit 
cannot be used in this case and adopting $b = 6.5$\kms, we find $\lnniii \le 14.03$
(instead of 13.55 in the thin limit) or [Ni/H] $\le -1.35$. 
%
\subsection{Temperature of the 21~cm absorbing gas}
%
When both \lya\ and 21~cm absorptions are detected, useful constraints
can be derived on the spin temperature of the gas, $T_\mathrm{s}$. 
The absorbers toward PKS~0454+039 and 3C~286 have been already discussed by 
Steidel et al. (1995) and Cohen et al. (1994). In the former, 21~cm absorption 
has
not been detected by Briggs \& Wolfe (1983) and therefore, only a lower 
limit on $T_\mathrm{s}$ could be inferred, $T_\mathrm{s} > 580$K 
(Steidel et al. 1995). The high resolution optical data recently obtained 
by Lu et al. (1996) can be used to get an even tighter constraint. Indeed, they 
find that the $b$ value for the main component is about 20\kms\ 
(see e.g. the unsaturated \mnii\ lines). The $b$ parameter relevant to the 
21~cm absorbing \hi\ cannot be larger which implies $T_\mathrm{s} > 870$K 
adopting a single temperature model. Using a similar assumption, Cohen et al. 
(1994) infer $T_\mathrm{s} \simeq 1200$~K for the DLAS in 3C~286.
Our result on $\nhi$ at $\za=0.437$ in 3C~196 cannot be used to 
constrain $T_\mathrm{s}$ for that absorber because the radio source 
is essentially extended (and then probes lines of sight distinct 
from the optical one). 

In PKS~1229$-$021, the situation is more favorable 
since a significant fraction of the flux at about 1 GHz (the frequency 
of the redshifted 21~cm line is 1018 MHz) originates from a compact 
component coincident with the optical quasar (see radio maps published 
by Kronberg et al. 1992). 
Following Brown \& Spencer (1979) we assume that 50\% of the 1GHz 
flux is emitted by the compact component
and that the latter is completely covered by the absorber 
(this corresponds to a size larger than 30pc). We then derive 
$T_\mathrm{s}$ = 170K. Part of the extended 
emission could also be covered by the 0.3950 absorber which would result in an 
increase of $T_\mathrm{s}$. However, such an effect is unlikely 
to significantly affect the previous estimate because i) the 21~cm line 
is narrow (FWHM $= 11$\kms )  
which suggests that the size of the absorbing region is much smaller than
that of a whole galaxy and ii) the absorber candidate (object
\#3 in Fig. 12 of paper I) does not cover the extended emission regions.
On the opposite, part of the \hi\ inducing the \lya\ 
absorption could be at relatively high temperature (e.g. $T_\mathrm{s} \ge $ 
1000K) 
and would then be inefficient in producing 21~cm absorption, 
which would imply an even lower $T_\mathrm{s}$
value for the rest of the gas (see the discussion by Wolfe et al. 1985). 
For instance, if 75\% of the gas is at
a temperature higher than 1000K, the remaining 25\% has to be at less than 49K.
We can therefore confidently conclude that, contrary to the DLAS in 
PKS~0454+039 and 3C~286 (see also Briggs \& Wolfe 1983), a significant fraction 
of the 
absorbing gas is at a low temperature, typical of \hi\ clouds
in the Galactic disk. 
\subsection{Neutral carbon}
\begin{figure}
\centerline{\psfig{figure=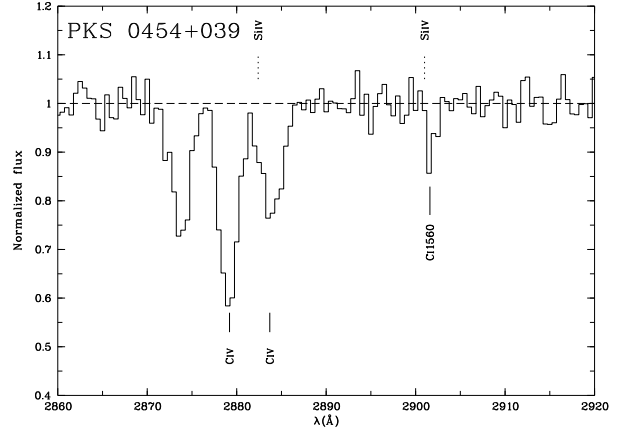,width=8cm,clip=t,angle=-90}}
\caption[]{\label{q0454ci} Portion of the G270H spectrum of PKS~0454+039 
comprising
the 1560 line from the DLAS. Dashed tick marks indicate the position 
expected for \siiv\
 lines from the weak \civ\  system at $\za = 1.0680$. The 
strong \civ\ doublet from the DLAS is also shown
}
\end{figure}

\begin{figure}
\centerline{\psfig{figure=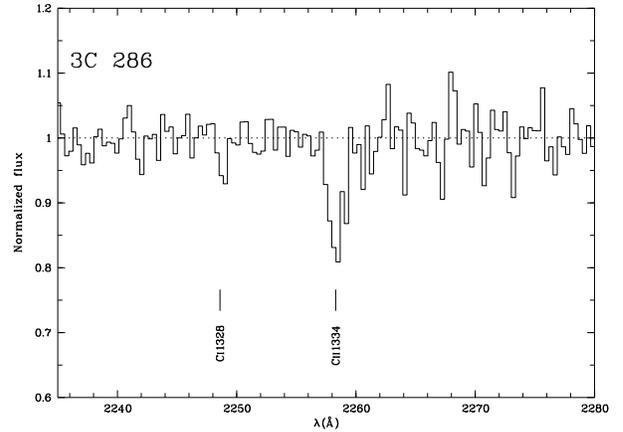,width=8cm,clip=t,angle=-90}}
\caption[]{\label{q3c286ci} Portion of the G190H normalized 
spectrum of 3C~286 comprising the \ci 1328 and \cii 1334 lines from the DLAS. 
Note the low noise level around 2250~\AA\, which corresponds to the top
of the QSO \lya\ emission line
}
\end{figure}

If physical conditions in the gas associated with the DLAS studied here were 
similar to those prevailing in the interstellar medium of our own Galaxy, 
neutral species should be present in detectable amounts. 
\ci\ especially can be searched for through its strong 1277, 1328, 1560 or 
1656 \AA\ transitions. The latter have been detected in some high redshift
DLAS (see e.g Blades et al. 1982; Ge et al. 1997) but, in several cases,
stringent upper limits have been obtained (Meyer \& Roth 1990; Black et al. 
1987). 
In order 
to investigate the presence of neutral gas, we made a specific search for 
\ci\ lines. For EX~0302$-$223 and PKS~1229$-$021, 
no useful constraint could be obtained because the features are expected 
in regions where either there are strong lines or $\sigma(\Wo)$ is too large.  
On the opposite, \ci1560 in PKS~0454+039 and \ci 1328 in 3C~286 are expected 
right
onto  one of the QSO emission line (\nv\ and \lya\ respectively) where the spectra 
are 
locally of excellent quality. 
In PKS~0454+039, we do see a weak line at 2901.55~\AA\ 
($\za = 0.8597$) with $\Wo = 0.14 \pm 0.05$ \AA\ (Fig.~\ref{q0454ci}).
An alternative identification could be \siiv 1402 (at $\za = 1.0684$) from the 
weak 
$\za = 1.0680$ metal system. Unfortunately, \siiv 1393 coincides with \civ\ from 
the 
DLAS and cannot be used to estimate the strength of \siiv 1402. The
wavelength match strongly favors an identification with \ci 1560 and we consider 
the latter as likely; higher resolution data are needed
to definitely establish the correct identification and the presence of
\ci . Similarly, in the G190H spectrum of 3C~286, 
there is a feature at 2248.87~\AA\ ($\za = 0.6924$) with $\Wo = 0.056 \pm$ 
0.022~\AA\  
(Fig.~\ref{q3c286ci}).
We consider the identification as certain because the line is also seen
in the G270H spectrum (although with a lower S/N) and because the wavelength 
match
is excellent. 
Assuming these lines to be optically thin 
we get $N(\mbox{\ci}) = 4.4 \times 10^{13}$ and $3.7 \times 10^{13}$\cm2\ for 
PKS~0454+039 and 3C~286 
respectively. In the second case, the quoted value is a lower limit because of 
possible
saturation effects (we get $5.0 \times 10^{13}$\cm2\ assuming instead $b=6.5$\kms). 

In order to
compare the physical conditions in these absorbers to those in our Galaxy we 
consider
the $\nhi - \nci$ plot given by Jenkins \& Shaya (1979). 
The DLAS in PKS~0454+039 appears close to that in Q~0013$-$004 (Ge \& Bechtold 
1997) and
to Galactic gas. On the other hand, the DLAS in 3C~286 is more like that in 
MC~1331+170 (Chaffee et al. 1988)
and PHL~957 (Black et al. 1987), i.e. significantly deficient in \ci\ with 
respect to Galactic 
gas.
However, given the low metal abundance seen in the absorber toward 3C~286, 
the inferred \ci/\hi\ ratio
suggests that physical conditions in the absorbers are relatively similar to those 
in our Galaxy and therefore, that there is enough dust to provide the
required shielding from UV photons with energy higher than 11.26 eV. 
\subsection{Dust grains}
Up to now, the evidence for dust associated with DLAS has been mostly 
statistical 
in nature, QSOs with DLAS showing in average steeper spectra than QSOs devoid of 
strong
systems (Pei et al. 1991). The overall pattern of metal abundances also strongly 
suggests 
that
selective depletion onto dust grains is effective in the absorbers (Pettini et 
al. 1997b; Kulkarni et al. 1997) 
although the interpretation of these data is still controversial (Lu et al. 
1996; Prochaska \& Wolfe 1996). The 2175~\AA\ feature would be a less 
ambiguous signature and can be
searched for in specific QSOs with DLAS. However, no clear detection has been 
obtained 
in any individual QSO (see e.g. Boiss\'e \& Bergeron 1988); this is 
generally 
taken as evidence for SMC or LMC-type extinction curves which display a less 
prominent 
feature. Recently, Malhotra (1997) found evidence for this feature in a composite 
spectrum
of QSOs with \mgii\ absorption. 

Regarding our targets, the 2175~\AA\ feature 
could be 
seen in PKS~0454+039 between the \civ\ and \ciii\ QSO emission lines (near 
4020~\AA)
in the excellent flux-calibrated spectrum obtained by Steidel \& Sargent (1992). 
A shallow depression is present centered about 40~\AA\ blueward of the expected
position and with a full width of about 200~\AA. 
Comparison with the composite spectrum computed by Zheng et al.
(1997) reveals that this feature is most likely intrinsic to the QSO.
In PKS~1229$-$021, it is expected at 3030~\AA\ near the end of our G270H 
spectrum: no
broad depression with an amplitude larger than 10\% is seen over a 300~\AA\ 
width 
interval. Finally, in 3C~286, some break is seen near 3680~\AA\ in the 
flux-calibrated spectrum presented by Aldcroft et al. (1994) which could 
be accounted for by a redshifted 2175~\AA\ feature with a depth of 15 to 20\%. 
The bluest part of the spectrum is noisy and probably affected by intrinsic 
broad
absorption; thus, the reality of that feature is difficult to assess.
The spectral index measured for PKS~1229$-$021 and 3C~286 between the 
\lya\ and \civ\ emission lines are 0.9 and 0.8 respectively which 
suggests little reddening. In the former, some bending is seen 
shortward
of the \ovi\ emission line but again, comparison with the composite spectrum of
Zheng et al. (1997) indicates an intrinsic origin.
\subsection{H$_2$ molecules}
H$_2$ and CO molecules have been searched for in the spectrum of QSOs with 
high redshift DLAS (see e.g. Black et al. 1987; Lanzetta et al. 1989). 
H$_2$ has been detected in two cases only: at $\za = 2.811$ in PKS~0528$-$250 
(Foltz et al. 1988) and recently at $\za = 1.9731$ in Q~0013$-$004 
(Ge \& Bechtold 1997). The former system is peculiar as it is at $\za
\simeq 
\ze$, so the latter case is the only clear detection of H$_2$ from gas which
is likely to be disk material. Aside from these two positive cases, low upper 
limits have been inferred for f, the fractional abundance of H$_2$ molecules
(typically $f \le 10^{-4}$ - $10^{-6}$; Black et al. 1987). One major difficulty 
encountered in these studies is that
H$_2$ lines are expected in the dense \lya\ forest where they can hardly 
be distinguished from \lya $-$only features. At lower redshift, the \lya\ forest
becomes less crowded and the situation is more favorable. 

Among the four QSOs
from our sample for which G190H or G270H spectra are available, three 
- EX~0302$-$223, PKS 0454+039
and 3C~286 - could display H$_2$ features (all the strong ones occur at
$\lambda_\mathrm{rest} \le 1120$~\AA). As emphasized by Black et al. (1987), 
anticoincidences are most significant and, in the spectrum of 
the three QSOs mentioned above, we have searched for windows 
which look free of any significant 
absorption and where strong H$_2$ lines are expected (Morton \& 
Dinerstein 1976; Foltz et al. 1988). Such regions can indeed be found 
(e.g. around 2035~\AA\ in PKS~0454+039 or around 1757~\AA\ and 1852~\AA\
in 3C~286: see Fig. 2 in Boiss\'e et al. 1998) 
which indicates that, at our detection limit, H$_2$ is not present.
The 3 $\sigma$ upper limit on $\Wr$ for unresolved H$_2$ lines in the three
QSOs is about 0.15 - 0.20~\AA. Unfortunately, this value cannot be translated 
easily into a limit on $N(\mathrm{H}_2)$ because the excitation 
temperature $T_{ex}$ and $b$ parameter are unknown. 
We can nevertheless obtain an upper limit by comparing the 
data to synthetic spectra computed by e.g. Foltz et al. (1988) and 
Lanzetta et al. (1989). For $b\simeq 5$ - $15$\kms\ and $T_\mathrm{ex}$ in the 
range
100 - 1000~K, $10^{18}$\cm2\ appears as a conservative upper limit on 
$N(\mathrm{H}_2)$.
Such a column density implies an upper limit on $f$ of $4. \times 10^{-3}$, 
$2. \times 10^{-3}$ and $7. \times 10^{-4}$ for EX~0302$-$223, 
PKS 0454+039 and 3C~286 respectively.
\subsection{Associated gas of high ionization}
Among the four DLAS studied at 1.5 - 2 \AA\ resolution, all have strong 
\civ\ - \siiv\ lines except that in 3C~286 (weak \civ, \siiv\ undetected). 
The \ovi\ doublet from the DLAS could have been
seen in PKS~0454+039 and 3C~286; it is present in the former absorber only, which 
displays an extensive range of ionization levels. 
\nv\ lines are 
undetected in EX~0302-223, PKS~1229-021 and 3C~286 while they are possibly 
present
in PKS~0454+039, blended with a group of \lya $-$only lines. We note that 
in the latter case, the line of sight to the QSO probes the halo of a  
compact galaxy.
The $3\sigma$ upper 
limits on $\Wr$ for undetected lines are in the range $0.15$ - $0.3$ \AA.
More data on \civ, \nv\ and \ovi\ absorption lines from low $z$ identified 
absorbers are needed to investigate the relation between the strength of 
high ionization features and the properties of the intervening galaxies. 
\section{Other metal-rich  and \lya\ forest absorbers}
\subsection{The dwarf galaxy near PKS~0454+039}
\begin{figure}
\centerline{\psfig{figure=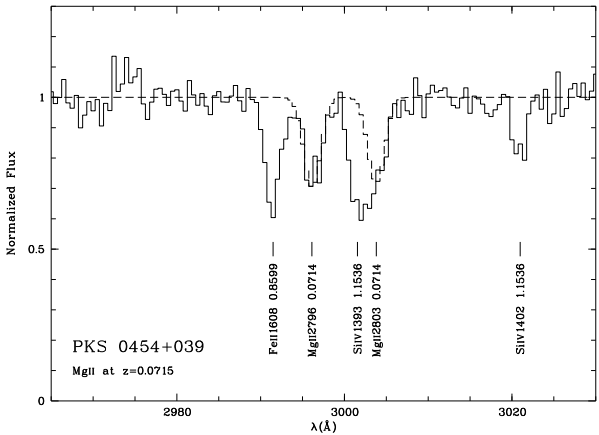,width=8cm,clip=t,angle=-90}}
\caption{\label{q0454mgii}Spectrum of PKS~0454+039 around the \mgii\ absorption 
induced by the $z=0.0718$ dwarf galaxy (full line). A fit to the $\za = 0.0715$
\mgii\ doublet is also shown (dashed line)}
\end{figure}

This galaxy, discovered by Steidel et al. (1993) has a redshift $z = 0.0718 
\pm 0.0004$,
and shows signs of star formation activity. Steidel et al. (1993) found no 
\caii\ absorption induced by this object in the spectrum of PKS~0454+039 
although 
the impact parameter, $D = 7.6\h50$~kpc, is relatively small. However, \caii\ 
does not probe efficiently the outer regions of galaxies where Ca is expected to 
be mostly in the form of Ca\,{\sc iii}. In Fig.~\ref{q0454mgii}, we display 
an enlarged portion of the G270H spectrum which shows unambiguously 
the presence of \mgii\ absorption from this dwarf galaxy. 
The 2796~\AA\ line appears at $\za=0.0715$, just in between the \feii 1608 line 
from the DLAS and the \siiv 1396 line at $\za=1.1537$. The latter feature is 
severely blended with the second \mgii\ doublet line. Absorption from the 
\feii 2600 line is also definitely present; the other major 
\feii\ transitions (at 2344 and 2382~\AA) are expected where strong lines 
from other systems occur.

In order to better probe the physical conditions in this absorber we searched 
for 21~cm absorption. Observations were performed in absentee at the ARECIBO
radiotelescope by M. Lewis and his collaborators (on the 21st and 22nd of
October 1993). The velocity range covered was 20700 - 22700\kms\ which 
corresponds to a redshift range of 0.0690 - 0.0757; the channel width was 
2.2\kms.   To check the pointing, 
a test galaxy was observed and detected. The continuum flux was measured 
to be 390.7 mJy (four different calibration sources have been used which gave
concordant results). No significant absorption was detected;
 the 3$\sigma$ upper limit is 2.4 mJy. Assuming that the velocity width of the 
\hi\ gas in front of the radiosource, $\Delta V$, is larger than the channel 
width,  we get the following upper limit on $\nhi$:\\
$\nhi \leq 1.1 \times 10^{18} \Delta V (km~s^{-1}) T_s/(100K)$\cm2.\\
This limit holds for a linear scale of $0.7 \h50$~kpc or below since the 
radiosource appears unresolved with a beam of 0.35\arcsec\ (Neff et al. 1989). 
Our result indicates that the intervening dwarf galaxy near PKS~0454+039 does 
not have a very
extended \hi\ disk which is consistent with \hi\ distributions observed in other 
nearby dwarf galaxies (Carignan \& Puche 1989; Jobin \& Carignan 1989). 

On the other hand, the detection of \mgii\ and \feii\ absorption shows that, in this 
class of objects, ionized gas is present well beyond the optical disk. 
Therefore, a large total mass is not a necessary condition for a galaxy
to display an extended ionized gas component, at least when star formation is 
occurring at a sufficiently high rate, as is the case in this galaxy. 
It should be noted that the impact parameter of this low redshift dwarf galaxy  
is 2.5 times smaller than the radius expected from the size - luminosity 
scaling law measured for brighter metal-rich absorbers 
(Guillemin \& Bergeron 1997; the absolute magnitude is M$_\mathrm{AB}$(B)=$-$17.02).
This is also the only dwarf galaxy 
giving rise to \mgii\ absorption in the available absorber samples 
although such objects could have
been detected up to redshifts of 0.2 - 0.3 (Steidel et al. 1994b;  Guillemin \& 
Bergeron 1997).

\subsection{Other cases}
Three other galaxies close to the QSO lines of sight studied in this paper have 
measured redshifts. We have already mentioned the bright spiral near 
EX~0302$-$223 
which gives rise to \mgii\ and \feii\ absorption at 0.118. We note that \feii 2600
is stronger than \mgii 2796 which suggests a large $\nhi$ value.
On the other hand, the relative strengths of \feii\  lines suggest that \feii 
2374 and \feii 2586 are nearly optically thin with $\lnfeii \simeq 14.2$ which 
rules
out a value of $\lnhi$ above 19 (the latter would correspond to [Fe/H] $\simeq
0$ assuming that most of the Fe is in the form of \feii ). There is however
no contradiction since, as discussed by Bergeron et al. (1994), the 
strong evolution in redshift of the diffuse UV background  implies that, at 
z $< 0.5$, large values of the \feii 2600/\mgii 2796 equivalent width ratio  
can be found in systems with  $\nhi\sim 10^{17}$ - $10^{18}$ cm$^{-2}$.

Another dwarf galaxy at low redshift and intermediate impact parameter, 
z$_{gal}$ = 0.199 and $D = 38.5\h50$~kpc, is present in the field 
of  PKS~1229$-$021 (G2 in Bergeron \& Boiss\'e 1991; G4 in Steidel et al. 1994a). 
In the spectrum of PKS~1229$-$021 several strong lines lie in the region where 
the \civ\ doublet is expected; the latter is not detected at a $3\sigma$ limit 
of $\Wr \le 0.30$~\AA\ for individual lines. The expected \aliii 1854 
line coincides with another strong absorption feature, but \aliii 1862 
could have been detected and is not present at a $3\sigma$ level of 0.15 \AA. 
This suggests that this dwarf galaxy  does not have  an extended gaseous halo.
This is not unexpected from its luminosity, M$_\mathrm{AB}$(B)=$-$18.0, since 
it lies just above the radius - luminosity scaling law of brighter 
metal-rich absorbers (Guillemin \& Bergeron 1997).
 
A low redshift, z$_{gal} = 0.2011$, bright, $L \sim L^{\star}$, galaxy  is  
present  
in the field of PKS~0454+039 (Steidel et al. 1993). As expected from its 
large impact parameter,  $D = 118\h50$~kpc, there is neither associated absorption
from  \mgii\  (Steidel et al. 1993) nor from \civ\ at a $3\sigma$
upper limit of $\Wr \le 0.30$~\AA.

\section{Discussion}

One first goal of our study was to observe the \lya\ line from our candidate 
high $\nhi$ systems. All of them turn out to have $\nhi$ larger than 
$2. \times 10^{20}$\cm2, which confirms the validity of the \feii /\mgii\ 
criterion used to select some of the systems (in EX~0302$-$223 and Q~1209+107). 
\subsection{Metal abundances}
The metal abundance estimates discussed in Sect. 4.1 have been 
gathered in Table~\ref{abond}. 
For all systems, relative abundances are given using our own 
determination 
of $\nhi$. The \znii\ and \crii\ column densities given by Meyer \& York (1992) 
and 
Steidel et al. (1995) have been adjusted according to the 
$f$ values used in this paper (the \nfeii\ derived by these authors from a 
curve of growth analysis have been adopted although they rely in part on $f$ 
values 
for \feii 2249 and \feii 2260 which were larger than the revised ones; these 
\nfeii\
may then be slightly underestimated, by 0.1 to 0.2 dex). Thus, our 
abundance estimates can be 
directly compared to those given by Lu et al. (1996) and Pettini et al. (1997a).
The uncertainty is typically 0.1 - 0.2 dex except for 3C~196 and Q~1209+107 
which 
have a less accurate $\nhi$ measurement. As is customary, relative abundances 
have been computed assuming $\nhii << \nhi$ and $\nfeii \simeq N$(Fe), 
etc (except for Ca which may be partly in the form of \caiii).

\begin{table*}
\caption{\label{abond} Relative metal abundances in damped \lya\ systems at $z 
<1$ 
([X/H] except for \caii\ for which $\lncaii$ is given)
}
\smallskip
\begin{tabular}{llllrrrrll}
\hline\noalign{\smallskip}
QSO            &  $\za$  &  $\nhi$        &   Si   &   Cr   &   Mn   &   Fe   &
   Ni   &    Zn   & \caii \\
\hline\noalign{\smallskip}
EX~0302$-$223  &  1.0095 & $20.39\pm0.04$ &        & $-0.9$ &$<-1.3$ &$>-1.4$ &
 $<-0.9$ & $-0.5$ &      \\
PKS~0454+039   &  0.8596 & $20.69\pm0.02$ &        & $-0.9$ & $-1.4$ & $-1.2$ & 
$-1.5$  & $-0.8$ &      \\
3C~196	       &  0.4368 & $20.8 \pm0.2 $ &        &        & $-0.9$ & $<0.3$ &  
       &        & 12.8 \\
Q~1209+107     &  0.6295 & $20.2 \pm0.1 $ &        &        &$<-0.3$ & $-0.9$ &  
       &        &      \\
PKS~1229$-$021 &  0.3950 & $20.75\pm0.07$ &$>-0.8$ &$<-0.8$ & $-0.8$ &$<-1.3$ 
&$-1.3$ & $-0.5$ & 12.3 \\
3C~286	       &  0.6922 & $21.19\pm0.02$ &        & $-1.6$ & $<-2.1$& $-1.8$ & 
 $<-1.3$& $-1.1$ & 12.5 \\
\noalign{\smallskip}\hline
\end{tabular}
\end{table*}
We first compare the pattern of relative abundances observed at high redshift by 
Lu et al. (1996) (see also Prochaska \& Wolfe 1996) 
and discussed by Pettini et al. (1997b)
and Kulkarni et al. (1997) to that of low redshift absorbers.  The 
estimates reported in Table~\ref{abond} appear in rough agreement 
with the compilation of [X/Zn] and [X/Fe] values presented by 
Kulkarni et al. (1997) for high redshift systems. 
Although covering a broad range, values for [Mn/H] and [Ni/H] are
roughly centered on the medians found at higher redshift,
$\langle$[Mn/Zn]$\rangle \simeq \langle$[Ni/Zn]$\rangle = - 0.65$ 
(Kulkarni et al. 1997), whereas the values (including one upper limit)
derived for [Fe/Zn] are all lower than that at higher redshift 
$\langle$[Mn/Zn]$\rangle = - 0.3$. Then, to first order, 
depletion onto dust grains 
seems to be effective in low redshift absorbers as it is at high 
redshift (Pettini et al. 1994; Pettini et al. 1997b;
Kulkarni et al. 1997); since in our Galaxy Zn is only slightly depleted, 
we shall follow Pettini et al. (1994) and use [Zn/H] as a
metallicity indicator. The [Fe/H] value in PKS~1229$-$021 is atypical in the 
sense that the upper limit inferred from our data appears very 
low when compared to [Mn/H] or, to a lesser extent, [Ni/H]. 
We note that in a recent study, Vladilo et al. (1997) find a
[Mn/Fe] ratio of 0.4 in a $\za = 0.558$ candidate DLAS
while our results on PKS~1229$-$021 imply [Mn/Fe] $\ge 0.5$;
both values appear high when compared to the high z estimates 
([Mn/Fe] $< 0.0$). However, given the difficulties quoted 
above for the measurement 
of [Fe/H] in the PKS~1229$-$021 DLAS, any conclusion (regarding e.g. the intrinsic 
nucleosynthetic pattern involved) would be premature and we stress that 
complementary observations (e.g. of \feii2367 or 
\feii2374) would be very valuable.

In the two cases where bright intervening spiral galaxies are present 
(systems in 3C~196 and Q~1209+107), 
the available Mn and Fe measurements suggest a 
relatively high metallicity. Assuming that the
abundance of these elements relative to Zn is similar to that at high $z$,
one gets estimates for [Zn/H] of $-$0.2 and $-$0.6 in 3C~196 and Q~1209+107 
respectively. For the former, the value estimated for [Zn/H] could be 
higher by 0.6 dex if the \nhi\ value given by Cohen et al. (1996)
had been adopted instead of ours. Similarly, for the latter, [Zn/H] could also be
higher by about 0.3 dex as it was derived using [Fe/H] (see above and Table 6).
Using the conservative estimates of [Zn/H] given above 
together with the measurement in PKS~1229$-$021 ([Zn/H] $= 
-$0.5), we find that the new low $z$ absorbers studied in this paper have 
[Zn/H] $\simeq -0.5$. Pettini et al. (1997a) have plotted all available 
measurements of [Zn/H] as a function 
of redshift (their Fig. 3). In this diagram, the three values quoted above  
indicate that systems with higher metallicities are present at lower redshift,
as expected from cosmic chemical evolution (Pei \& Fall 1995). 
Pettini et al. (1997a) also
present a binned version of this plot (their Fig. 4); 
in the low redshift bin which includes four systems at
$\za \le 1.5$, the column density weighted average 
metallicity is $\langle$[Zn/H]$\rangle = -0.98$ at  $\langle \za 
\rangle \simeq 1$. Including our three new 
estimates yields a higher value, $\langle$[Zn/H]$\rangle = -0.70$ at 
$\langle \za \rangle = 0.77$. A few other DLAS have been measured recently. 
In 3C~336, Steidel et al. (1997) find [Fe/H] $= -1.2$ and 
$\nhi = 2. \times 10^{20}$ \cm2 at $\za = 0.656$. 
Two more DLAS have also been discovered in the spectra of the bright 
QSOs HE~1122$-$1649 (at $\za = 0.68$) and HE~0515$-$4414 (at $\za = 1.15$)
with values for [Zn/H] of $< -1.52$ and $-0.85$ and for $\nhi$ of 
$3.1 \times 10^{20}$ 
\cm2\ and $2.0 \times 10^{20}$ \cm2\ respectively (de la Varga \& Reimers 1998). 
We checked that inclusion of the latter does not modify the 
above value for  $\langle$[Zn/H]$\rangle$ 
(this is because they involve low \hi\ column densities and because
it is appropriate to consider the \nhi\ weighted average).

To summarize, we find that extending the sample at lower redshifts brings 
$\langle$[Zn/H]$\rangle$
somewhat closer to the prediction of a metallicity 
approaching Solar values as $\za$
goes to zero (Pei \& Fall 1995), and further, that a large 
scatter is present, as at high redshift. The latter is likely to reflect the 
large variety of the observed absorber morphologies (paper I) and
the spread in impact parameters (Phillips \& Edmunds 1996).
\subsection{The bias induced by dust and physical conditions in the absorbers}
Several authors have considered that the extinction induced by dust within 
intervening galaxies may affect the statistics of distant QSOs 
(Ostriker \& Heisler 1984; Boiss\'e \& Bergeron 1988; Fall \& Pei 1993). Fall \& 
Pei (1993)
stressed that the same bias also affects the statistics of DLAS themselves, 
especially of those with the largest $\nhi$ and developed a method to 
correct for these effects.
Boiss\'e (1994, 1995) pointed out that, because dust is closely linked to 
metallicity
and to the formation of molecules, extinction probably results in a preferential
selection of QSOs with systems displaying a low metallicity and H$_2$ content. 

It is noteworthy that the dust obscuration bias is expected to have 
the strongest effect at z $\simeq$ 1, precisely in our range of interest 
(Fall \& Pei 1995; their Fig. 5); this is 
because metallicity decreases while the extinction per dust grain in the 
observer's frame increases (due to the rising extinction law) when going at 
higher redshift. Indeed, in the present study, we find that the two faintest QSOs (which
could not be observed with the HST at a resolution appropriate to detect metal
lines) display systems with relatively high metallicities. We also find no very 
large $\nhi$ values (e.g. $\ge 10^{22}$ \cm2) while this could have been expected 
given the 
low impact parameters involved (e.g. in 3C~196). As discussed by Boiss\'e (1995), 
the absence of such systems which would be quite easily recognized in the numerous 
low resolution optical spectra obtained so far, further supports the reality
of a cut-off caused by extinction. 

\begin{figure}
\centerline{\psfig{figure=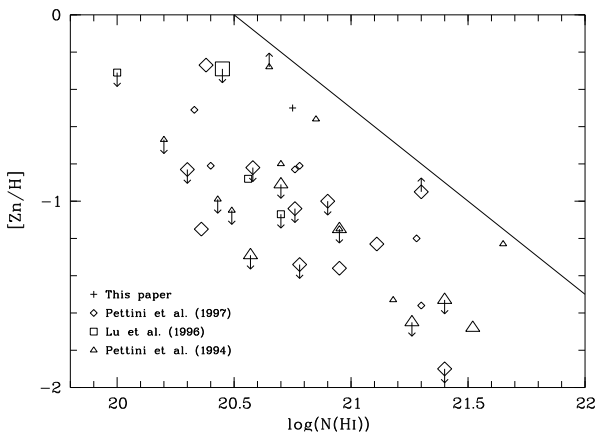,width=8cm,clip=t,angle=-90}}
\caption{\label{zn_h-nhi}[Zn/H] versus $\lnhi $ for 37 damped \lya\ systems; 
small symbols correspond to $\za \le 2.15$ and large symbols to $\za \ge 2.15$. 
The line in the upper right corresponds to $\nznii = 1.4 \times 10^{13}$\cm2 or 
to Galactic material inducing $A_\mathrm{V} \simeq 0.27$ (see text)}
\end{figure}

These results alone are only indicative but after 
the extensive work by Pettini et al. (1994, 1997a) and Lu et al. (1996), there 
now exists a
reasonably large sample of systems with $\nhi$ and [Zn/H] measurements 
(again, since Zn does not deplete onto 
grains, it is a good indicator of the amount of metals available to form dust)
and effects of the extinction bias might become apparent in the data themselves. 
We then plot [Zn/H] as a function of \hi\ column density, since the latter
should determine to first order the strength of the effect
(Fig.~\ref{zn_h-nhi}; the $\za > \ze$ system in PKS~0528$-$250 has not been 
included). We note 
a very clear deficiency of systems having at the same time a large $\nhi$ {\it 
and} 
a high metallicity (the absence of data points in the bottom left part is
just due to observational limitations). 
The available measurements appear to be distributed in the diagram
as if the amount of metals along the line of sight, as estimated from N(Zn),
were constrained to
be less than about $1.4 \times 10^{13}$\cm2, which corresponds to the line drawn in 
Fig.~\ref{zn_h-nhi}. This 
result is all the more striking as, in an unbiased sample, 
one would have rather expected the opposite trend since the large $\nhi$ values 
should 
correspond to the innermost parts of galaxies where the metallicity is 
presumably 
higher. To verify
that the evolution of [Zn/H] in redshift does not affect our analysis (through 
a 
possible variation of $<\nhi>$ with $\za$), we consider separately measurements 
at 
$\za < 2.15$ (the median of $\za$ values)
and $\za > 2.15$; no tendency is seen for high $z$ systems
to cluster at large $\nhi$. 

One alternative explanation would involve the opacity of Zn lines; in the upper 
right part of
Fig.~\ref{zn_h-nhi}, \nznii\ is large and the optically thin approximation could 
no longer be
valid which would result in an underestimate of [Zn/H]. 
While this may be true
in a few cases (two measurements in Fig.~\ref{zn_h-nhi} are in fact given as 
lower limits) the high resolution data presented by 
Lu et al. (1996) indicates that the whole set of [Zn/H] values is unlikely
to be affected by saturation effects. 
We then conclude that dust extinction is effective at inducing 
a 
preferential selection of QSOs with systems having a low metal content. 

The
upper bound on N(Zn) present in Fig.~\ref{zn_h-nhi} can be interpreted as an 
upper bound on A$_V$ of about 0.3 for Galactic-type material; the latter is
assumed to be characterized by [Zn/H] $\simeq -0.2$ (Sembach et al. 1995) and 
we have adopted the $\nhi$/A$_V$ ratio given by Bohlin et al. (1978).
For larger values of A$_V$, significant effects are indeed expected to occur 
since in the 
rest frame of the absorber, it is the UV part of the extinction curve that is 
involved.
It is often argued that, given the low dust-to-gas ratio inferred from 
the observed systems 
(Pei et al. 1991), the effects of extinction are necessarily small. 
Indeed, if extinction effects were estimated assuming the {\it average} 
dust-to-gas ratio inferred from the reddening of QSOs with DLAS, one would 
predict 
negligible effects in Fig.~\ref{zn_h-nhi}.
However, this reasoning does not take into account the properties of the 
systems that are missed and which are very likely to be those with the largest 
dust-to-gas ratio. 
Fall \& Pei (1993) precisely pointed out that the (unknown) dispersion 
in the dust-to-gas ratio is an important limitation in our ability to correct for 
the effects of this bias. Metal abundances in DLAS (as in nearby galaxies) show a 
large 
scatter at any redshift, and similarly large variations can be expected for the 
dust-to-gas ratio itself, including the possibility that galaxies with 
dust-to-gas 
ratio and metallicities even larger than Galactic exist, since we have no reason 
to believe that the Milky Way is particularly dust and metal rich. 
Fig.~\ref{zn_h-nhi}
confirms that in the available samples, a selection with respect to 
absorbing gas properties is present.

A direct consequence of the bias induced by dust extinction is that estimates
of $\Omega_\mathrm{g}$, the total amount of gas (see e.g. Wolfe et al. 1995;
Storrie-Lombardi et al. 1996), are necessarily highly uncertain. First, the 
relative 
contribution to $\Omega_\mathrm{g}$ of large $\nhi$ systems is important
and these are precisely the most heavily biased. Second, H$_2$, which is not 
taken into 
account, may represent a non-negligible fraction of the total gas mass. In the 
local Universe, this fraction is estimated to be in the range 30 - 50 \% 
(Casoli et al. 1997). The evolution in redshift of this 
quantity is unknown but we already observe that by $z \simeq $ 1 - 2, galaxies form 
stars at 
a high rate which strongly suggests the presence of a large H$_2$ mass. This is 
also indicated by CO observations of distant galaxies (Solomon et al. 1992; 
Omont
et al. 1996; Alloin et al. 1997) which probably give a lower limit to the true 
amount of H$_2$ because the CO to H$_2$ conversion factor is likely to be 
higher due to a lower metallicity. Note however that, to first order, the
uncertainty on the H$_2$ amount does not 
affect the method developed by Fall \& Pei (1993) since their estimate
of extinctions is based primarily on the column density of metals, which 
is an observed quantity. 

As noted above, the extinction bias is expected to
have large effects at $\za \simeq 1$. This is also true at $\za < 1$
because the UV spectroscopic observations 
required to study these 
systems are very demanding in terms of QSO brightness. This tends to reduce the 
evolution of [Zn/H] (which further helps to understand the discrepancy discussed 
by Pettini et al. 1997a between the observed
low $z$ bin and model predictions) and might also mimic an evolution of the shape 
of the $\nhi$ 
distribution, such as that observed by Wolfe et al. (1995).

Since molecules and dust are closely related, the extinction bias is likely to 
affect 
the apparent H$_2$ content as well. One point that has been overlooked when 
discussing 
absorption line studies is that such
methods yield very little information on the densest phases of the interstellar 
medium
where stars form. This is for two reasons. First, the surface coverage factor of 
the latter
is quite small and therefore the probability to intersect intervening material 
of this type
is very low. Second, any background QSO will be strongly dimmed by the 
associated extinction and will remain undetected or appear too faint for 
spectroscopic 
studies; this effect is reinforced by the marked tendency of molecules to 
assemble in 
dense opaque clouds, due to self-shielding. 
For galaxies at $z \le$ 1, we have ample evidence for the presence of dense 
and dusty molecular clouds like those in our own Galaxy (Wiklind \& Combes
1996; 
Casoli et al. 1996). Nevertheless, when searching for dust and molecules 
from low $z$ systems in QSO spectra, we get essentially the same null result 
as at higher redshift! 

Therefore, one must not draw any definite conclusion 
from the 
observed low amount of molecules and lack of clear signatures from dust grains 
in DLAS. Only the observation of fainter QSOs 
would lead to a more representative view of the 
interstellar medium
in these distant galaxies (Boiss\'e 1994, 1995). 
Even if H$_2$ molecules have not been detected in our QSO sample, 
the likely detection of \ci\ in two systems 
suggests that physical conditions (radiation field, density, ...) 
are not very different from those in the Solar
neighborhood, at least in some absorbers. A similar conclusion is reached by 
Ge \& Bechtold (1997) for gas at $\za = 1.97$.

\subsection{Relation between damped absorber morphology and intervening gas 
properties}
The galaxies proposed in paper I as the absorbers have not yet been confirmed 
spectroscopically. This will be a difficult task because in ground-based 
observations
a large fraction of the QSO light is superimposed onto the galaxy 
emission and a redshift can be 
measured only if the latter shows prominent spectral signatures. Additional
problems arise from the detection of several metal systems, each of them 
implying
the presence of a galaxy close to the QSO line of sight. Since we now have a 
detailed census of metal systems in four out of six QSOs from our sample, let us 
summarize 
the status of the proposed identifications and discuss their reliability:\\
	- EX~0302$-$223: four objects are present at low impact parameter (all 
with a
relatively compact morphology). Given its lower impact parameter, galaxy \#2 is 
the best
candidate. The other three galaxies, of which two are very close and probably
form an interacting system, could give rise to the metal-rich systems at 
$\za = 0.9109$ and 1.3284. The $\za = 0.4196$ absorber is identified with 
a very bright galaxy at large impact parameter (Guillemin \& Bergeron 1997).\\
	- PKS~0454+039: there is little ambiguity in this case since only one 
object is seen near the QSO. The galaxy is very compact and the QSO line 
of sight probes regions which are outside the stellar component.\\
	- 3C~196: the large spiral is very likely to be at 0.437 because its 
luminosity would have to be extremely large if it were at 0.87.
This very extended galaxy does not show any [{O\,{\sc ii}}]3727 emission
but the \caii\ absorption doublet has been tentatively identified; in galaxy
\# 3, strong emission lines at a redshift close to that of the QSO have been
detected (Drouet d'Aubigny \& Bergeron, in preparation). However, the latter
may not be the $z = 0.871$ absorber, since the intervening gas does not cover
the whole broad line region and should thus be  very close to the active nucleus 
(Cohen et al. 1996),\\
	- Q~1209+107: the proposed identification with galaxy \#2 is reliable 
because no other candidate is present close to the line of sight.
Furthermore, its color is consistent with a 
spiral galaxy at $z \simeq 0.6$,\\
	- PKS~1229$-$021: although we could reject objects \# 2, \#4 and \#6 as 
potential absorbers since they are associated with knots in the QSO radio jet, 
two objects (a compact galaxy, \#3, and a low surface brightness galaxy, \#5) 
could be at 0.3950. Galaxy \#3 is favored because of its lower impact
parameter,\\
	- 3C~286: three fairly bright objects (\#2a, \#2b and \#2c) have been
detected at less than 1 arcsec from the QSO line of sight. Al three could 
contribute to the DLAS. Alternatively, since \#2a and \#2b are located roughly 
symmetrically to the QSO, they could be part of the QSO's host galaxy, \#2c
being then the absorber.

Some of these identifications could be erroneous if, as in the case of 3C336
(Steidel et al. 1997), none of the galaxies close to the QSO line of sight 
turned out to be at the DLAS redshift. However, from a statistical point of 
view,
the presence of galaxies at low impact parameter in all the studied cases cannot be
due to chance and very likely, most of the proposed identifications are correct.

Among the gas properties that may be connected to absorber morphology
let us consider metallicity and kinematics.
We already noted that the two cases with intervening
spiral galaxies correspond to relatively high metallicities. The low
[Zn/H] value measured in 3C 286 in spite of the small impact parameter 
is consistent with the proposed absorber of fairly low surface brightness. 
This case may then be similar to the DLAS studied by Steidel et al. 
(1997): no galaxy could be detected despite an intensive search which 
suggests a faint object centered on the QSO. The two compact absorbers 
display intermediate metallicities; these galaxies are reminiscent of the 
blue nucleated galaxies that appear at $z \simeq $ 0.5 in galaxy surveys
(Schade et al. 1996). The most intriguing case is the DLAS
in PKS~1229$-$021: although the amorphous aspect suggests an unevolved 
object, the Zn and Mn abundances are clearly higher than in the two 
previous cases, with a possibly unusual abundance pattern.

The absorber in PKS~1229$-$021 is also remarkable for the kinematics of the
absorbing gas. The asymmetrical velocity distribution revealed by the data 
obtained
by Lanzetta \& Bowen (1992) is the prototype of what Prochaska \& Wolfe (1997) 
consider
to be the signature of a massive disk. However, at first sight, 
such an interpretation
of absorption profiles is not supported by the appearance of the 
damped absorber in 
HST images as discussed by Pettini et al. (1997a). Recent numerical simulations 
also confirm (at least for the high $z$ DLAS) that the presence of an edge-leading 
asymmetry is not an unambiguous signature of a rapidly rotating disk (Haehnelt
et al. 1997). We also note that in the other case of a proposed low surface 
brightness galaxy
(in 3C 286), the kinematics seems to be very different since all or most of the 
absorption arises from the single narrow component which induces 
the 21cm absorption line
(Meyer \& York, 1992). 

Detailed information on the kinematics is also available 
for the damped \lya\ system in PKS~0454+039 (Lu et al. 1996). Absorption is 
spread more or less continuously over 140 \kms\ with four maxima and no 
characteristic asymmetry. No simple picture emerges from the yet very scarce 
systems for which the required data are available. A larger sample is necessary 
to check if the absorption profiles are related to the large-scale 
kinematics of the intervening galaxy itself (as proposed by Prochaska \& 
Wolfe 1997) or rather are governed by e.g. the recent star formation 
activity (through supernovae explosions). In the present context, the two 
systems
induced by bright spiral galaxies (in 3C 196 and Q~1209+107) are of much 
interest and would clearly deserve high resolution spectroscopic observations.
\subsection{Perspectives}
The results obtained in our study and the above discussion clearly show the 
need for complementary 
observations. Obviously, it is very desirable to confirm spectroscopically the
candidate absorbers proposed in paper I, especially when several galaxies are 
present at low impact parameter. Then, it is important to improve 
the accuracy of some measurements presented in this paper and 
extend the set of metal elements considered in order
to compare in more details the abundance patterns seen at low and high z. 
In particular, it is important to confirm the high metallicity suggested for the 
systems caused by intervening spiral galaxies in the spectra 
of 3C 196 and Q~1209+107. For the two latter cases, 
the absorber properties are very well 
characterized by HST images; high resolution optical spectra would provide
the velocity distribution of the gas and thus help to establish the 
phenomena governing the kinematics of the absorbing material. 

Moreover, the observed broad range of absorber morphologies and absorbing 
gas properties indicate that a larger sample is needed to get a correct overall 
view of the origin of low $z$ damped \lya\ systems. Possibly, some types of
absorbers have not yet been detected; the recent study by Lanzetta et al.
(1997) shows that early-type galaxies do also contribute to the damped \lya\
absorber population. Presently, only two or three absorbers 
of each type have been found 
and it is not possible to assess the relative contribution of the various 
classes 
of galaxies. Observations of several absorbers from the same class,
but probed at different impact parameters and inclinations, would be crucial to 
understand the relation between the absorbing gas and the parent 
galaxy properties.

Finally, as discussed in Sect. 6.2, we strongly suspect that surveys of 
damped \lya\ samples drawn from the observation of fainter QSOs
would reveal more systems with large \nhi\  
and high metallicity, whatever the redshift. However, the most reddened 
QSOs would not be observable in their UV rest-frame and the intervening
gas properties should then be derived from observations at longer wavelengths.
Nevertheless, searches for damped \lya\ absorption in QSOs fainter by two or 
three magnitudes could reveal absorption systems in which the presence of 
dust grains and molecules
is much more conspicuous than in the systems investigated so far, and thus
would allow a more comprehensive study of the evolution of the 
interstellar medium in galaxies.
Spectroscopic surveys of fainter QSOs should then not only enlarge the present
samples but also lead to a more representative view of the whole diversity of 
the damped \lya\ absorber population.

\acknowledgements{We are grateful to Chuck Steidel and Patrick Petitjean for 
allowing us to use some of their data on PKS~1229$-$021, and also to
Murray Lewis and 
his collaborators who observed for us PKS~0454+039 with the ARECIBO radiotelescope. 
We also wish to thank Mike Fall for insightful comments on an earlier version of this 
paper and Ross Cohen, Fran\c{c}oise Combes, Max Pettini and Evelyne Roueff 
for helpful discussions on various aspects of this work.}

\end{document}